\documentclass[preprint,floatfix] {revtex4} 
\newcommand{\rvec}{\mathrm {\mathbf {r}}} 
 
\usepackage{graphicx}
\usepackage{tikz}
\usepackage{subfigure}
\usepackage{xcolor}
\usepackage{amsmath}
\usepackage{enumerate}
\usepackage{tabularx}
\usepackage{nicefrac}
\usepackage{enumitem}
\usepackage{color, soul}
\definecolor{darkblue}{rgb}{0,0,0.5}
\setulcolor{darkblue}
\usepackage{multirow}
\usepackage{longtable}

\begin{document}

\title{Degeneracy and metallic character in free and confined weakly coupled plasmas: with and without electric field}

\author{Neetik Mukherjee}
\email{pchem.neetik@gmail.com.}

\author{Amlan K.~Roy}
\altaffiliation{Corresponding author. Email: akroy@iiserkol.ac.in, akroy6k@gmail.com.}

\affiliation{Department of Chemical Sciences, 
IISER Kolkata, Mohanpur-741246, Nadia, WB, India}

\begin{abstract}
Incidental degeneracy and metallic character is probed for weakly coupled plasmas in free and confined environments. The generality of incidental 
degeneracy in quantum mechanical systems is discussed and demonstrated. It is a fundamental property of free and confined quantum systems. In plasmas, 
at a given $n, \ell$state there exists $\frac{(n-\ell)(n-\ell+1)}{2}$ number of incidental degenerate states. Such degeneracy condition involves shell 
confinement model, where a particleis encaged inside two concentric sphere. Apart from that, Dipole oscillator strength and polarizability are examined 
in free and confined conditions for ground and some low-lying $\ell$ states. In excited states, negative $\alpha^{(1)}$ is recorded. Further, metallic 
behavior of H-like plasmas is investigated. The impact of external static electric field on these degeneracy, dipole OS, dipole polarizability are 
examined with utmost interest.  Pilot calculation are done with, (i) Debye plasmas and, (ii) Exponential screened coulomb potentials employing the 
Generalized pseudo-spectral (GPS) method. 

\vspace{2mm}
\end{abstract}
\maketitle
\section{Introduction}
Hydrogen atom represents the simplest two-body system consisting of a proton and an electron. The Schr\"odinger equation (SE) 
offers exact analytical closed-form solution in both non-relativistic and relativistic domain. Probing hydrogenic systems is 
extremely important, as they can act as a precursor to understand the quantum effects in more complex structure \cite{solyu12, 
paul09}. In the visible universe, hydrogen constitutes more than 90$\%$ of all atoms and contributes three quarter of its mass 
\cite{paul09}. Further, it is a well known fact that, in the inner core of the Sun, hydrogen fuses together to form helium and 
release energy. This is the source of energy in our solar system. Further, its abundance in the interiors of 
Jupiter and Saturn is well known. In these planets, it is present in both metallic and liquid form, which is responsible for their origin of 
magnetosphere \cite{stevenson74,guillot04,sudbo04,gregoryanz20}. To be precise, most of the celestial bodies ranging from giant planets to 
recently discovered brown dwarfs \cite{horn91} are primarily composed of dense hydrogen and helium plasmas. In recent years the investigation of atomic 
processes in plasma environment has become a subject of topical interest. It elicits the screening effect of plasma on atoms embedded in such situations. 

High energy-density physics covers a broad range of plasmas from very hot to dense conditions. In this scenario, there occurs a coupling 
between plasma electrons and immersed atoms, leading to a change in electronic properties. The composite influences of plasma free-electron 
density $(n_{e})$ and temperature $(T)$ plays a crucial role in stabilizing a bound state by controlling the strength of 
this coupling. The coupling parameter $(\Gamma)$ is defined as \cite{murillo04,das14,mukherjee21},
\begin{equation}
\Gamma =\frac{r_{-}}{a} = \frac{E_{\mathrm{coulomb}}}{E_{\mathrm{thermal}}}=\frac{Q^{2}}{aT}, 
\end{equation}              
where $r_{-}$ is the critical radius of a circular volume beyond which no plasma electron can approach the atom \cite{murillo04}, 
$a=\left(\frac{3}{4\pi n_{e}}\right)^{\frac{1}{3}}$ is the inter-particle spacing, called \emph{ion-sphere radius}, $Q$ denotes 
the charge of ion. $\Gamma << 1$ signifies high $T$, low $n_{e}$ condition (weakly coupled), while, $\Gamma >> 1$ represents low $T$, high 
$n_{e}$ situation (strongly coupled). In a given plasma, the correlated many-particle interactions are described by an average screening 
potential, incorporating the collective effects due to the presence of charged cloud \cite{flores18}.    

In a dense plasma, the competing effect of $n_{e}$ and $T$ are modulated through Debye length, $\lambda_{D}=\left(\frac{T}{4\pi Q^{2}n_{e}}\right)^{\frac{1}{2}}$. 
At high $T$ and low $n_{e}$, $\lambda_{D}$ possesses higher value, and as a consequence, offers greater number of bound states. On the contrary, 
$\lambda_{D}$ decreases at low $T$ and high $n_{e}$, leading to a decrease in the count of bound states. Further, the plasma-tail effect (emerges due to the 
existence of asymptotic part in potential) appears in the picture with rise in $T$. The simplest plasma condition is introduced by invoking a 
prototypical Debye-H\"uckel potential (DP), with the form: $V_{1}(r)=-\frac{Z}{r}e^{-\frac{r}{\lambda_{D}}}$ ($Z$ is nuclear 
charge). In last two decades, DP was investigated vigorously with utmost attention. The effect of plasma screening on the energy spectrum 
\cite{solyu12,paul09a,bahar14,bahar16}, virial and Hellmann-Feynman theorems \cite{montgomery18}, two proton transitions \cite{paul08,paul09}, 
transition probabilities connecting electron-impact excitation \cite{jung95,jung96,song03}, inelastic electron-ion scattering \cite{gutierrez94, 
yoon96}, Fisher information, Shannon entropy, statistical complexities \cite{zan17} etc., have been studied. Numerical values of critical 
screening constant ($\lambda^{(c)}$) for ground and low-lying excited states were presented in \cite{stubbins93}. Recently, an empirical relation 
between $\lambda^{(c)}_{n,\ell}$ and $Z$ was proposed in \cite{mukherjee21}. The dynamic plasma screening on DP was investigated in 
\cite{jung97,qi09b,liu08,liu08a}. The relativistic correction on plasma screening was also considered \cite{poszwa15}. Influence of external 
static electric field on energy spectrum of DP is also studied \cite{bahar14}. Spectroscopic properties together with multipole oscillator 
strength (OS) and static multipole polarizabilities were evaluated for H-like atoms under the influence of DP \cite{saha02,qi08,qi09,qi09a,
bassi12} using various numerical methods. A time-dependent variation perturbation approach was adopted to calculate transition probability, OS, 
static dipole polarizability for ground state at varied $\lambda_{D}$ values \cite{saha02a}. Recently, a generalized pseudo spectral (GPS) 
method is employed to give high-quality results of OS and polarizabilities in ground and excited states \cite{zhu20}. Several well known methods 
like, integration based shooting technique \cite{das12}, linear variation method \cite{kang13}, numerical symplectic integration method 
\cite{qi08,qi09,qi09a}, mean excitation energy based approximation formula \cite{bassi12}, etc., were also invoked to extract these properties. 
The hyperpolarizability of H atom under spherically confined DP was estimated in \cite{saha11}. Further, polarizabilities are also computed for 
confined DP plus ring-shaped potentials \cite{yadav21}. In the above references, however, calculations were mostly limited to ground state. 
In a recent work, multipole ($k=1-4$) OS, and subsequent polarizabilities are reported for $1s,2s$ states of \emph{confined} DP \cite{mukherjee21}.    

The exponential cosine screened Coulomb potential (ECSCP), is expressed as, $V_{2}(r)=-\frac{Z}{r}e^{-\frac{r}{\lambda_{D}}} \cos(\frac{r} {\lambda_{D}})$, 
exerts a stronger effect compared to DP. The existence of oscillatory part manifests in a combined screening and wake effect around a slow-moving 
test charge in high $n_e$, low $T$ plasma. The cosine term drives the quantum force enacting on plasma electrons to predominate over statistical 
pressure exerted by plasmas \cite{shukla08,jiao21}. In quantum plasma, $\lambda_{D}$ depends on the wave number of electron. Its eigenvalue and 
eigenfunctions were studied in a number of methods, \emph{viz.}, perturbation and variation \cite{lam72}, Pad\'e scheme 
\cite{lai82}, shooting \cite{singh83}, SUSY perturbation \cite{dutt86}, asymptotic iteration \cite{bayrak07}, Laguerre polynomial \cite{lin10}, 
variation using hydrogenic wave functions \cite{paul11}, J-matrix \cite{nasser11}, GPS \cite{roy13}, basis expansion method with Slater-type 
orbitals \cite{lai13}, symplectic integration \cite{qi16}, etc. The influence of $\lambda_{D}$ on energy spectrum \cite{bahar14,bahar16}, 
photoionization cross-section \cite{lin10,lin11}, electron-impact excitation \cite{song03}, etc., has been probed as well.  
Similar to DP, attempt was made to determine the characteristic $\lambda_{D}$ beyond which, the bound state ceases to exist \cite{diaz91,
mukherjee21}. The laser-induced excitation on confined H atom (CHA) in ECSCP was reported in terms of laser pulse, $r_c, \lambda_D$ using 
Bernstein-polynomial \cite{lumb14}. Variations of $f^{(1)}, \alpha^{(1)}$ against $\lambda_{D}$ were reported in \cite{singh83,dutt86,qi16,lai13,roy16,
jiao21}. But, like DP, ECSCP in \emph{confined environment} has not been investigated so far in a sufficiently thorough fashion. Only some limited  
recent works on $f^{(k)}, \alpha^{(k)} (k=1-4)$ for $1s,2s$ states \cite{mukherjee21} and $\alpha^{(1)}$ for confined ECSCP in a ring-shaped 
potential \cite{yadav21} are available. 

The concept of confinement introduces several astonishing change in chemical and physical properties in a given quantum system. This happens 
due to the rearrangement of atomic orbitals in such scenario \cite{grochala07,roy15}. In this endeavour, it is pertinent to discuss behavior of 
hydrogen under stressed condition. Because under such situation (which can arise due to external high pressure) it can exhibit metallic 
character \cite{howie12,howie15,gregoryanz20,dias17}, which triggers superconductivity and superfluidity at certain optimum $T$ and $P$ 
\cite{ginzburg04,mcmohan12}. Actually, several decades ago, in 1935, it was proposed that molecular hydrogen would become an atomic metal at 
certain a characteristic high pressure \cite{wigner35}. In the new millennium, discovery and development of modern experimental techniques 
have made it possible to investigate these phenomena \cite{dias17,howie12}. Raman and visible transition spectroscopy reveals that, at high 
pressure ($200 \mathrm{Gpa} -315 \mathrm{Gpa}$), several phase transitions occur leading to new state of hydrogen having metallic character 
\cite{sudbo04,howie12,knudson15}. With an increase in external pressure, conductivity increases and resistivity decreases \cite{weir96}. Moreover, 
in 1968, it was proposed that, metallic modification of hydrogen will guide us to high-temperature superconductor \cite{ashcroft68}. In a 
recent work, this feature was observed experimentally in room temperature under $325 \ \mathrm{Gpa}$ \cite{simpson16}. In theoretical perspective, 
metallic behavior of H-atom in ground state was in reported in \cite{sen02} by employing the \emph{Herzfeld criterion} for insulator $\rightarrow$ 
metal conversion \cite{herzfeld27}. Lately, further exploration has revealed that, in H atom, this occurs only in case of $\ell=0$ orbitals 
\cite{mukherjee21a}. In chemistry, the shell confinement condition can be illustrated by citing the examples of trapping of an atom/molecule 
within \emph{metal organic framework} \cite{efros16,fei21}, inside fullerene cage and zeolite cavity \cite{nascimento11}. The sintering effect gets 
minimised; as a consequence the catalytic activity and thermal stability of certain \emph{noble} metals get improved 
\cite{peng18,rao18,kumar19,lai21,fei21}. Further, such condition amplifies photoluminescence character in nano crystals by reducing non-radiative 
Auger processes \cite{efros16,fan20} and dispels defects in polymer crystals \cite{shi13,khadilkar18} etc. Shell confinement plays key role in 
energy storage \cite{kumar18,shuang19,wang20} and therapeutics \cite{hastings19} and pollution control \cite{qin20,zhang20}. After having such 
multipurpose applications, shell confinement model has rarely been studied. Hence the literature is very scarce. 

We know that, \emph{incidental degeneracy} is achieved when the energy of confined state equals that of an unconfined state. For this to happen,  
the impenetrable boundary should be placed at the position of \emph{radial nodes} of unconfined state \cite{sen05,mukherjee21a}. Such degeneracy 
introduces a \emph{shell-confined} model, where a H atom was confined within two concentric spheres of inner ($R_{a}$) outer ($R_{b}$) radii. In 
this context, our primary objective is to pursue \emph{incidental degeneracy and metallic character} in two prototypical plasmas, namely (i) 
DP and ii) ECSCP. Towards this goal, we consider $f^{(k)}$ and $\alpha^{(k)}$ ($k=1$) in ground and excited states. The dependency of incidental degeneracy 
upon principal ($n$) and orbital ($\ell$) quantum numbers is critically explored. It directs us to evaluate the correct number of such degenerate 
states related with a given plasma energy in free condition. Moreover, we also examine the impact of external electric field on this  
degeneracy and metallic behavior (via $\alpha^{(1)}$) in hydrogenic plasma. Pilot calculations are done by invoking the GPS method. To the best of 
our knowledge, most of the results are reported here for first time. The article is constructed in following parts: Sec.~II illustrate a brief 
description about the formalism used in present work. We discuss about the origin of incidental degeneracy in Section~III. Section~IV portrays 
a thorough discussion of results. Finally, we conclude with a few remarks in Sec.~V.

\section{Theoretical formalism}
The radial Schr\"odinger equation (SE) for a single particle spherically confined system is given (atomic unit employed unless otherwise stated) as,
\begin{equation}\label{eq:1}
\left[-\frac{1}{2} \ \frac{d^2}{dr^2} + \frac{\ell (\ell+1)} {2r^2} + V_{e}(r) + V_{c}(r)\right] \psi_{n,\ell}(r)=\mathcal{E}_{n,\ell}\ \psi_{n,\ell}(r), 
\end{equation}
Whereas, $V_{e}=Fr\cos\theta$ illustrates the \emph{external electric field}. In field free cases, $F=0$ or $\theta=\frac{\pi}{2}$. In present case, 
calculations are done considering $\theta$ as parameter. Therefore, the Overall SE in spherical polar coordinate can be separable into radial and 
angular part.    

Further, $V_{c}$ is the desired confined potential given below \cite{sen05,sen02}, 
\begin{equation}\label{eq:2}
\begin{aligned}    
V_{c}(r)=\begin{cases}
=v(r) \ \  \mathrm{for}   \ \ \ R_{a} \le r \le R_{b} \\
=\infty \ \ \ \mathrm{for} \ \ \ 0 \le r \le R_{a} \\
=\infty \ \ \ \mathrm{for} \ \ \ r \ge R_{b}.    
\end{cases}
\end{aligned}
\end{equation}

$v(r)$ is the generalised plasma potential, which is expressed in the following form,
\begin{equation}
v(r)=-\frac{Z}{r}(1+br)e^{\left(-\frac{r}{\lambda}\right)}\cos\left(c~\frac{r}{\lambda}\right),
\end{equation}
where $b,c$ are real numbers. 
\begin{enumerate}
\item $b=0, c=0$, represents DP with the form,
\begin{equation}
v(r)=-\frac{Z}{r}e^{\left(-\frac{r}{\lambda_{1}}\right)}.
\end{equation}
\item $b=0,c\ne 0$, demonstrates ECSCP, which is expressed as,   
\begin{equation}
v(r)=-\frac{Z}{r}e^{\left(-\frac{r}{\lambda_{2}}\right)}cos{\left(c~\frac{r}{\lambda_{2}}\right)}.
\end{equation}
For calculation purpose $c=1$.  
\end{enumerate}
Here, $Z$ is the nuclear charge. Here, $V_{c}(r)$ will be referred as generalised confined condition (GCS). Depending upon the values 
of $R_{a},R_{b}$, four distinct conditions can be envisaged, they are,  
\begin{enumerate}
\item
When $R_{a}=0$, $R_{b}=\infty$ $\rightarrow$ \emph{free condition} (FC). 
\item
When $R_{a}=0$, $R_{b}= r_{c}$, a finite number $\rightarrow$ \emph{confined situation} (CS).
\item
When $R_{a} \ne 0$, $R_{b} \ne \infty$, with $R_{a}, R_{b}$ finite, $\rightarrow$ \emph{shell confined condition} (SCC).
\item
If $R_{a} \ne 0$, $R_{b}=\infty$, $\rightarrow$ \emph{left confined situation} (LCS).
\end{enumerate}     
In order to estimate the energy, spectroscopic properties the GPS method was invoked. Over the time, it has been successfully 
used to calculate several bound-state properties of various central potentials \cite{roy14,roy14a,roy14b,mukherjee18,mukherjee19,mukherjee20, majumdar20, majumdar21}.
 
\subsection{Multipole polarizability}
The static multipole polarizability can be conveniently written as,  
\begin{equation}\label{eq:3}
\alpha^{(k)}_{i}=\alpha^{(k)}_{i}(\mathrm{bound})+\alpha^{k}_{i}(\mathrm{continuum}).
\end{equation} 
Conventionally $\alpha^{(k)}_{i}$ is expressed in sum-over states form \cite{das12}. However it can also be directly 
estimated by adopting the standard perturbation theory\cite{dalgarno62}. Eq.~(\ref{eq:4}) modifies to,
\begin{equation}\label{eq:4}
\begin{aligned}
\alpha^{(k)}_{i} & = \sum_{n}\frac{f^{(k)}_{ni}}{(\mathcal{E}_{n}-\mathcal{E}_{i})^{2}}
-c\int\frac{|\langle R_{i}|r^{k}Y_{kq}(\rvec)|R_{\epsilon p}\rangle|^{2}}{(\mathcal{E}_{\epsilon p}-\mathcal{E}_{i})} \ \mathrm{d}\epsilon, \\
\alpha^{(k)}_{i}(\mathrm{bound}) & = \sum_{n}\frac{f^{(k)}_{ni}}{(\Delta \mathcal{E}_{ni})^{2}}, \ \ \ \ \ 
\alpha^{k}_{i}(\mathrm{continuum}) = c\int\frac{|\langle R_{i}|r^{k}Y_{kq}(\rvec)|R_{\epsilon p}\rangle|^{2}}{(\mathcal{E}_{\epsilon p}-\mathcal{E}_{i})} \  \mathrm{d}\epsilon.
\end{aligned}
\end{equation}  
In Eq.~(\ref{eq:4}), the first and second terms indicate bound and continuum contributions respectively, $f^{(k)}_{ni}$ is the 
multipole oscillator strength ($k$ is a positive integer), $c$ is a \emph{real} constant depends only on $\ell$ quantum number. Here, $f^{(k)}_{ni}$ is normally expressed as, 
\begin{equation}\label{eq:5}
f^{(k)}_{ni}=\frac{8\pi}{(2k+1)}\Delta\mathcal{E}_{ni}|\langle r^{k} Y_{kq}(\rvec)\rangle|^{2}. 
\end{equation}    
Illustrating the initial ($|n \ell m\rangle$$|n \ell m\rangle$) and final ($|n^{\prime}\ell^{\prime}m^{\prime}\rangle$) states one can easily derive,  
\begin{equation}\label{eq:6}
f^{(k)}_{ni}=\frac{8\pi}{(2k+1)} \ \Delta\mathcal{E}_{ni} \ \frac{1}{2\ell+1}\sum_{m}\sum_{m^{\prime}} |\langle n^{\prime}\ell^{\prime}m^{\prime}|r^{k}Y_{kq}(\rvec)|n \ell m \rangle|^{2}.
\end{equation}
Wigner-Eckart theorem and sum rule for \emph{3j} symbol further leads to,  
\begin{equation}\label{eq:7}
f^{(k)}_{ni}=2 \ \frac{(2\ell^{\prime}+1)}{(2k+1)} \ \Delta\mathcal{E}_{ni} \ |\langle r^{k}\rangle^{n^{\prime}\ell^{\prime}}_{n \ell}|^{2} \
\left\{\begin{array}{c} \ell^{\prime} \ \ k \ \ \ell\\ 0 \ \ 0 \ \ 0 \end{array}\right\}^{2}. 
\end{equation}
The transition matrix element is given by following radial integral,
\begin{equation}\label{eq:8}
\langle r^{k} \rangle = \int_{0}^{\infty} R_{n^{\prime} \ell^{\prime}}(r) r^{k} R_{n \ell} (r) r^{2} \mathrm{d}r.
\end{equation}
Note that, $f^{(k)}_{ni}$ depends only on $n, \ell$. We compute $f^{(1)}$, 
$\alpha^{(1)}$, for states with $\ell=0-2$. It is necessary to point out the multipole 
oscillator strength sum rule as, 
\begin{equation}\label{eq:9}
S^{(k)}=\sum_{m}f^{(k)}=k\langle \psi_{i}|r^{(2k-2)}|\psi_{i}\rangle, 
\end{equation}
where the summation incorporates all the bound and continuum states.

\subsubsection{Herzfeld criteria for Metallic Character}
According to Herzfeld criteria \cite{herzfeld27}, insulator to metallic conversion occurs after attaining a threshold volume having the form,
\begin{equation}\label{eq:10}
V=\frac{4}{3}\pi \alpha_{d},	
\end{equation}
where $\alpha_{d}$ is the dipole polarizability of the atom or molecule. Let us assume that, an atom or a molecule is trapped inside a concentric 
sphere of radius $r_{c}$, then, Eq.~(\ref{eq:10}) modifies to the form,
\begin{equation}\label{eq:11}
\begin{aligned}
V =\frac{4}{3} \pi r_{c}^{3} & = \frac{4}{3}\pi \alpha_{d}  \\
r_{c}^{3} & =\alpha_{d}.
\end{aligned}
\end{equation}
Further, considering that, the particle is confined inside a spherical shell having inner and outer radius $R_{a}$ and $R_{b}$ respectively. Therefore,
Eq.~(\ref{eq:10}) becomes,
\begin{equation}\label{eq:12}
\begin{aligned}
V  =\frac{4}{3} \pi \left(R_{b}^{3}-R_{a}^{3}\right) & = \frac{4}{3}\pi \alpha_{d}  \\
\left(R_{b}^{3}-R_{a}^{3}\right) & =\alpha_{d}.
\end{aligned}
\end{equation}
It is important to point out that, at $R_{a}=0$ Eq.~(\ref{eq:12}) reduces to Eq.~(\ref{eq:11}), therefore $R_{b}=r_{c}$. However, in H-like atoms 
it has been found that, for a arbitrary $n,\ell$ state $\alpha_{d}$ decreases with reduction in $r_{c}$ \cite{mukherjee21}. As a matter of fact 
$V > \alpha_{d}$, hence metallic character is not seen. On the contrary, $\alpha_{d}$ increases with decrease in $\Delta R= \left(R_{b}-R_{a}\right)$. 
In a given $(n,0)$ state, for each of these $R_{b}$ values there exist a characteristics $R_{a}$ values after which 
$\alpha_{d}>\left(R_{b}^{3}-R_{a}^{3}\right)$, as a results metallic character is seen \cite{mukherjee21a}. In the present endeavour, we aim to verify 
this idea for plasma systems. 

\section{Origin of incidental degeneracy}
In quantum mechanics a \emph{free particle} of mass $a$ is allowed to move in all direction in space with energy $k^{2}/2a$ ($k$ is a real number).
Therefore, a free particle possesses any $+$ve value of energy. The introduction of the boundary condition $\psi_{n}(x)=0$ at $x=L_{1}$ and $x=L_{2}$
($L_{2} > L_{1}$) leads us to most prototypical system in quantum mechanics: \emph{particle in a box} (PIB). Application of such boundary condition
invokes both quantization and confinement together in the form of an impenetrable box of length $L=(L_{2}-L_{1})$ and energy $\frac{n^{2}\pi^{2}}{2aL^{2}}$.
The eigenfunction of the $n$th state is expressed as $\psi_{n}(x)=\sqrt{\frac{2}{L}} \sin \left(\frac{n\pi}{L}\right)x$. 

Let us consider a PIB with length $L_{m}$, whose $m$th state has the same energy with the $n$th state of another box with length $L_{n}$. Thus, one
can write,
\begin{equation}\label{eq:13}
\begin{aligned}
\frac{n^{2}\pi^{2}}{2aL_{n}^{2}} & =\frac{m^{2}\pi^{2}}{2aL_{m}^{2}}  \\
\frac{n^{2}}{L_{n}^{2}} & =\frac{m^{2}}{L_{m}^{2}}  \\
L_{m} & =\left(\frac{m}{n}\right)L_{n}.
\end{aligned}
\end{equation}

In PIB, a node at the point $x_{j}$ appears when, $\psi_{n}(x_{j})=\sqrt{\frac{2}{L_{n}}} \sin \left(\frac{n\pi}{L_{n}}\right)x_{j}=0$. Therefore,
\begin{equation}\label{eq:14}
\begin{aligned}
\sin \left(\frac{n\pi}{L_{n}}\right)x_{j} & =0, \\
\mathrm{thus}, \ \ \ \ x_{j} & = \left(\frac{m}{n}\right)L_{n}, \\
\mathrm{finally}, \ \ \ \ L_{m} & =\left(\frac{m}{n}\right)L_{n}.  
\end{aligned}  
\end{equation}

Therefore, from Eqs.~(\ref{eq:13}), and (\ref{eq:14}) we can conclude that, when $n > m$ cases, $L_{m}$ ($m=1,2,3,..$) are the nodal points of the 
$n$th state of the PIB having length $L_{n}$. Now, placing an impenetrable boundary at certain $L_{m}$ value one can approach to isoenergic states 
with energy $\frac{n^{2}\pi^{2}}{2aL_{n}^{2}}$. For, example, if $m=1$, then $n$ such state appears, similarly, for $m=2$, we get, $(n-1)$ states. 
Obeying the same norm for $m=m$ state provides $(n-m+1)$ number of states. Therefore, in completeness, for a given $n$ we get $\frac{n(n+1)}{2}$ 
number of incidental degenerate states. It is noteworthy to mention that, when $m > n$, then for each of the $m$ values there arise an isoenergic 
state with energy $\frac{n^{2}\pi^{2}}{2aL_{n}^{2}}$ and box length $L_{m}$. But, they are not originated due to the incidental degeneracy condition, 
and for the sake of lucidness we call then additional degenerate states. 

\begingroup           
\squeezetable
\begin{table}
\caption{Incidental degeneracy in 1-D QHO associated with $n=1,2$.}
\centering
\begin{ruledtabular}
\begin{tabular}{l|llllll}
Serial & No. of nodes & state ($m$) & $R_{a}$  & $R_{b}$ & Energy & 1-D QHO state ($n$)\\
\hline
1a      & 0           &  0                          & $-\infty$  & 0         & $\frac{3\alpha}{2}$ & 1  \\
1b      & 0           &  0                          & 0          & $\infty$ & $\frac{3\alpha}{2}$ & 1  \\
1c      & 1           &  1                          & $-\infty$  & $\infty$  & $\frac{3\alpha}{2}$ & 1  \\
2a      & 0           &  0                          & $-\infty$  & $-\frac{1}{2\sqrt{\pi}}$ & $\frac{5\alpha}{2}$ & 2  \\
2b      & 0           &  0                          & $-\frac{1}{2\sqrt{\pi}}$  & $\frac{1}{2\sqrt{\pi}}$ & $\frac{5\alpha}{2}$ & 2  \\
2c      & 0           &  0                          & $\frac{1}{2\sqrt{\pi}}$ & $\infty$ & $\frac{5\alpha}{2}$ & 2  \\
2d      & 1           &  1                          & $-\infty$  & $\frac{1}{2\sqrt{\pi}}$ & $\frac{5\alpha}{2}$ & 2  \\
2e      & 1           &  1                          & $-\frac{1}{2\sqrt{\pi}}$ & $\infty$  & $\frac{5\alpha}{2}$ & 2  \\
2f      & 2           &  2                          & $-\infty$  & $\infty$ & $\frac{5\alpha}{2}$ & 2  \\
\end{tabular}
\end{ruledtabular}
\end{table}  
\endgroup  
Thus, depending upon the value of $m$, two different types of degeneracy appear in PIB. Now, the point is to find out the answer, whether these degeneracies 
are appearing either due to quantization or due to confinement. In this context to determine the origin of these degeneracies, we invoke 1-Dimensional quantum 
harmonic oscillator (1-D QHO) having the form, $v(x)=\frac{1}{2}\alpha^{2}x^{2}$ ($\alpha=2\pi \nu$) and $\mathrm{E}_{n}=(n+\frac{1}{2})\alpha$. Detailed 
analysis reveals that, in 1D QHO incidental degeneracy exists but not the accidental degeneracy. Thus, replacing the infinite barrier by a continuous potentialwould not hamper the incidental degeneracy (qualitative manner) in the system. Therefore, one can conjecture that, incidental degeneracy is a fundamental property of quantum systems and a consequence of quantization. However, additional degeneracy arises purely due to confinement. The total analysis in 1-D QHO is demonstrated in Table~I involving $n=1,2$ states ($R_{a}, R_{b}$ are the nodal points of 1-D QHO. These are used to invoke shell confined condition.)

\squeezetable          
\begin{longtable}{p{0.8cm}|p{0.8cm}|p{0.8cm}p{1.1cm}|ccc|ccc}
\caption{Incidental degeneracy in DP and ECSCP, for $\lambda_{1},\lambda_2=200.0$ a.u.} \\
\hline
\multirow{2}{*}{\parbox{0.8cm}{Serial No.}} & \multirow{2}{*}{\parbox{0.8cm}{Ref. state}} & \multirow{2}{*}{\parbox{0.8cm}{Deg. state}} & \multirow{2}{*}{\parbox{0.9cm}{No. of nodes}} 
& \multicolumn{3}{c|}{DP} & \multicolumn{3}{c}{ECSCP} \\
\cline{5-10} 
&  &  &   & $R_{a}$  & $R_{b}$ & Energy & $R_{a}$  & $R_{b}$ & Energy   \\
\hline
\endfirsthead
\multicolumn{10}{r}%
{\tablename\ \thetable\ -- \textit{Continued from previous page}} \\
\hline
\multirow{2}{*}{\parbox{0.8cm}{Serial No.}} & \multirow{2}{*}{\parbox{0.8cm}{Ref. state}} & \multirow{2}{*}{\parbox{0.8cm}{Deg. state}} & \multirow{2}{*}{\parbox{0.9cm}{No. of nodes}} 
& \multicolumn{3}{c|}{DP} & \multicolumn{3}{c}{ECSCP} \\
\cline{5-10} 
&  &  &   & $R_{a}$  & $R_{b}$ & Energy & $R_{a}$  & $R_{b}$ & Energy   \\
\hline
\endhead
\hline \multicolumn{10}{r}{\textit{Continued on next page}} \\
\endfoot
\hline
\endlastfoot
$2as$ &  $2s$   &  $1s$ & 0 & 0            & 2.00009870    & $-$0.12007414 & 0             & 2.0000026110   & $-$0.12000171     \\
$2bs$ &  $2s$   &  $1s$ & 0 & 2.00009870   &  100          & $-$0.12007414 & 2.0000026110  &  100           & $-$0.12000171     \\
$2cs$ &  $2s$   &  $2s$ & 1 & 0            & $\infty$      & $-$0.12007414 & 0             &  $\infty$      & $-$0.12000171     \\
$3as$ &  $3s$   &  $1s$ & 0 & 0            & 1.902122031   & $-$0.05072018 & 0             &  1.901934382   & $-$0.05056383     \\
$3bs$ &  $3s$   &  $1s$ & 0 & 1.9021220310 & 7.1008210     & $-$0.05072018 & 1.901934382   & 7.09825460     & $-$0.05056383     \\
$3cs$ &  $3s$   &  $1s$ & 0 & 7.1008210    &  100          & $-$0.05072018 & 7.09825460    &  100           & $-$0.05056383     \\
$3ds$ &  $3s$   &  $2s$ & 1 & 0            & 7.1008210     & $-$0.05072018 &  0            &  7.09825460    & $-$0.05056383     \\
$3es$ &  $3s$   &  $2s$ & 1 & 1.9021220310 &  100          & $-$0.05072018 & 1.901934382   &   100          & $-$0.05056383     \\
$3fs$ &  $3s$   &  $3s$ & 2 & 0            & $\infty$      & $-$0.05072018 &  0            &  $\infty$      & $-$0.05056383     \\
$4as$ &  $4s$   &  $1s$ & 0 & 0            & 1.8719797570  & $-$0.02653747 & 0             & 1.87167478     & $-$0.02627512     \\
$4bs$ &  $4s$   &  $1s$ & 0 & 1.8719797570 & 6.614980540   & $-$0.02653747 & 1.87167478    & 6.611235230    & $-$0.02627512     \\
$4cs$ &  $4s$   &  $1s$ & 0 & 6.614980540  & 15.54031890   & $-$0.02653747 & 6.611235230   & 15.52028550    & $-$0.02627512     \\
$4ds$ &  $4s$   &  $1s$ & 0 & 15.54031890  & 100           & $-$0.02653747 &  15.52028550  &  100           & $-$0.02627512     \\
$4es$ &  $4s$   &  $2s$ & 1 & 0            & 6.614980540   & $-$0.02653747 & 0             & 6.611235230    & $-$0.02627512     \\
$4fs$ &  $4s$   &  $2s$ & 1 & 1.8719797570 & 15.54031890   & $-$0.02653747 & 1.87167478    &  15.52028550   & $-$0.02627512     \\
$4gs$ &  $4s$   &  $2s$ & 1 & 6.614980540  &  100          & $-$0.02653747 & 6.611235230   &  100           & $-$0.02627512     \\
$4hs$ &  $4s$   &  $3s$ & 2 & 0            & 15.54031890   & $-$0.02653747 & 0             &  15.52028550   & $-$0.02627512     \\
$4is$ &  $4s$   &  $3s$ & 2 & 1.8719797570 & 100           & $-$0.02653747 & 1.87167478    &  100           & $-$0.02627512     \\
$4js$ &  $4s$   &  $4s$ & 3 &  0           & $\infty$      & $-$0.02653747 & 0             & $\infty$       & $-$0.02627512     \\
\hline
$3ap$ &  $3p$   &  $2p$ & 0 & 0            & 6.002624940   & $-$0.05070822  &  0             & 6.000155460    & $-$0.0505627661 \\
$3bp$ &  $3p$   &  $2p$ & 0 & 6.002624940  & 100           & $-$0.05070822  &  6.000155460   & 100            & $-$0.0505627661 \\
$3cp$ &  $3p$   &  $3p$ & 1 &  0           & $\infty$      & $-$0.05070822  &  0             &  $\infty$      & $-$0.0505627661 \\
$4ap$ &  $4p$   &  $2p$ & 0 &  0           & 5.531842190   & $-$0.02652592  &  0             & 5.5282437020   & $-$0.0262733156     \\
$4bp$ &  $4p$   &  $2p$ & 0 & 5.531842190  & 14.49475650   & $-$0.02652592  &  5.5282437020  & 14.4747240030  & $-$0.0262733156     \\
$4cp$ &  $4p$   &  $2p$ & 0 & 14.49475650  &  100          & $-$0.02652592  &  14.4747240030 &  100           & $-$0.0262733156     \\
$4dp$ &  $4p$   &  $3p$ & 1 & 0            & 14.49475650   & $-$0.02652592  & 0              &  14.4747240030 & $-$0.0262733156     \\
$4ep$ &  $4p$   &  $3p$ & 1 & 5.531842190  & 100           & $-$0.02652592  & 14.4747240030  &  100           & $-$0.0262733156     \\
$4fp$ &  $4p$   &  $4p$ & 2 & 0            & $\infty$      & $-$0.02652592  & 0              &  $\infty$      & $-$0.0262733156     \\
$5ap$ &  $5p$   &  $2p$ & 0 & 0            & 5.35879755    & $-$0.015428340 & 0              & 5.3538548120   & $-$0.0150560898     \\
$5bp$ &  $5p$   &  $2p$ & 0 & 5.35879755   & 13.318341040  & $-$0.015428340 & 5.3538548120   & 13.293465580   & $-$0.0150560898     \\
$5cp$ &  $5p$   &  $2p$ & 0 & 13.318341040 & 26.4658350    & $-$0.015428340 & 13.293465580   &  26.378176830  & $-$0.0150560898     \\
$5dp$ &  $5p$   &  $2p$ & 0 & 26.4658350   &  100          & $-$0.015428340 & 26.378176830   & 150            & $-$0.0150560898     \\
$5ep$ &  $5p$   &  $3p$ & 1 & 0            & 13.318341040  & $-$0.015428340 & 0              & 13.293465580   & $-$0.0150560898     \\
$5fp$ &  $5p$   &  $3p$ & 1 & 5.35879755   & 26.4658350    & $-$0.015428340 & 5.3538548120   & 26.378176830   & $-$0.0150560898     \\
$5gp$ &  $5p$   &  $3p$ & 1 & 13.318341040 &  150          & $-$0.015428340 & 13.293465580   & 150            & $-$0.0150560898     \\
$5hp$ &  $5p$   &  $4p$ & 2 & 0            & 13.318341040  & $-$0.015428340 & 0              & 26.378176830   & $-$0.0150560898     \\
$5ip$ &  $5p$   &  $4p$ & 2 & 5.35879755   &  150          & $-$0.015428340 & 5.3538548120   &  150           & $-$0.0150560898     \\
$5jp$ &  $5p$   &  $5p$ & 3 & 0            & $\infty$      & $-$0.015428340 & 0              & $\infty$       & $-$0.0150560898     \\
\hline
$4ad$ &  $4d$   &  $3d$ & 0 & 0            & 12.020605190  & $-$0.0265028061 & 0             &  12.002149730  & $-$0.026269684     \\
$4bd$ &  $4d$   &  $3d$ & 0 & 12.020605190 & 100           & $-$0.0265028061 & 12.002149730  &  200           & $-$0.026269684     \\
$4cd$ &  $4d$   &  $4d$ & 1 & 0            &  $\infty$     & $-$0.0265028061 & 0             &  $\infty$      &  $-$0.026269684    \\
$5ad$ &  $5d$   &  $3d$ & 0 & 0            & 10.91249580   & $-$0.015406227  & 0             & 10.889614250   & $-$0.015050727     \\
$5bd$ &  $5d$   &  $3d$ & 0 & 10.91249580  & 24.2196030    & $-$0.015406227  & 10.889614250  & 24.13278320    & $-$0.015050727     \\
$5cd$ &  $5d$   &  $3d$ & 0 & 24.2196030   & 150           & $-$0.015406227  & 24.13278320   & 220            & $-$0.015050727     \\
$5dd$ &  $5d$   &  $4d$ & 1 & 0            & 24.2196030    & $-$0.015406227  & 0             & 24.13278320    & $-$0.015050727     \\
$5ed$ &  $5d$   &  $4d$ & 1 & 10.91249580  & 150           & $-$0.015406227  & 10.889614250  & 250            & $-$0.015050727     \\
$5fd$ &  $5d$   &  $5d$ & 2 & 0            & $\infty$      & $-$0.015406227  & 0             &  $\infty$      & $-$0.015050727     \\
$6ad$ &  $6d$   &  $3d$ & 0 & 0            & 10.46443610   & $-$0.00947425   &  0            &  10.436313970  & $-$0.008994213     \\
$6bd$ &  $6d$   &  $3d$ & 0 & 10.46443610  & 22.08628590   & $-$0.00947425   & 10.436313970  & 21.98896250    & $-$0.008994213     \\
$6cd$ &  $6d$   &  $3d$ & 0 & 22.08628590  & 39.9757030    & $-$0.00947425   & 21.98896250   & 39.70523610    &  $-$0.008994213    \\
$6dd$ &  $6d$   &  $3d$ & 0 & 39.9757030   &  125          & $-$0.00947425   & 39.70523610   &  250           & $-$0.008994213     \\
$6ed$ &  $6d$   &  $4d$ & 1 & 0            & 22.08628590   & $-$0.00947425   & 0             &  21.98896250   &  $-$0.008994213    \\
$6fd$ &  $6d$   &  $4d$ & 1 & 10.46443610  & 39.975703     & $-$0.00947425   & 10.436313970  & 39.70523610    &  $-$0.008994213    \\
$6gd$ &  $6d$   &  $4d$ & 1 & 22.08628590  &  150          & $-$0.00947425   & 21.98896250   &  250           &  $-$0.008994213    \\
$6hd$ &  $6d$   &  $5d$ & 2 & 0            & 22.08628590   & $-$0.00947425   & 0             &  21.98896250   & $-$0.008994213     \\
$6id$ &  $6d$   &  $5d$ & 2 & 10.46443610  &  125          & $-$0.00947425   & 10.436313970  &   250          &  $-$0.008994213    \\
$6kd$ &  $6d$   &  $6d$ & 3 & 0            &  $\infty$     & $-$0.00947425   &  0            &   $\infty$     &  $-$0.008994213    \\
\end{longtable} 

\section{Result and Discussion}
Here we mention all four (FC, CS, SCC, LCS) models under the general heading of GCS. However, since the energy characteristics of these 
models are quite different. Demonstrative results are reported for both DP and ECSCP ($Z=1$). However, 
analogous results can be found for $Z \ne 1$ cases. At the onset, we discuss several salient features of incidental
degeneracy obtained by placing the boundary at respective nodal positions of DP and ECSCP in both \emph{free} and \emph{confined} situations.
Next, we report $f^{(k)}(Z)$ and $\alpha^{(k)}(Z)$ ($k=1$) for some low-lying states of DP and ECSCP, involving FC, CS, SCC models. Further,
we have evaluated  $\alpha^{(1)}(Z)$ for $1s, 2s, 2p, 3s, 3d, 4s$ states. It is necessary
to point out that, in case of degeneracy, radial boundaries are opted specifically at the nodes of DP/ECSCP to demonstrate their
role. Whereas for $f^{(k)}, \alpha^{(k)}$, no such condition was taken into consideration. Further, it may be noted that, for either of the plasmas (DP and ECSCP), 
all these calculations are performed in the range of $\lambda$ between $0-200$ a.u. Here we report results involving $\lambda=200$ a.u. Because, at this
$\lambda$ values higher number of bound states exists in DP and ECSCP. Pilot calculations are executed in two steps. At first we investigate incidental
degeneracy and spectroscopic properties for DP and ECSCP. Later, we look into the effect of external electric field on incidental degeneracy, 
dipole oscillator strength and polarizability.    

\subsection{Incidental Degeneracy}
Following \cite{sen05}, it is a well known fact that, if $R_{a}, R_{b}$ of a \emph{Generalized confined hydrogen atom} (GCHA) \cite{mukherjee21a} 
marge with certain specific radial nodes of $(n,\ell)$ state of \emph{free H-atom} (FHA), as a result there exists $(n^{'}-\ell-1)$ number of nodes 
in between them. Moreover, energy of such a $(n^{'},\ell)$ GCHA state appears to be degenerate to that of a FHA state. Equation~(\ref{eq:2}) depicts that, 
\emph{shell confined} condition introduces four different systems. This degeneracy proposes the inter-connection amongst them. Here, we involve ourselves to 
investigate the same for plasma systems in both \emph{confined} and \emph{free} conditions, where the systems are devoid of \emph{accidental degeneracy}. 
Further, we would also determine the number of \emph{degenerate states} attached with a given plasma energy in either of the potentials. It should also
be keep in mind that, the conclusions for \emph{free} DP and ECSCP are drawn assuming the existence of infinite number of bound states in them. However, with 
decrease in bound states, number of such degenerate states reduces.

\squeezetable         
\begin{longtable}{p{0.8cm}|p{0.8cm}|p{0.8cm}p{1.1cm}|ccc|ccc}
\caption{Incidental degeneracy in \emph{confined} DP and ECSCP ($r_{c}=5$ a.u.), for $\lambda_{1},\lambda_2=200.0$ a.u.} \\
\hline
\multirow{2}{*}{\parbox{0.8cm}{Serial No.}} & \multirow{2}{*}{\parbox{0.8cm}{Ref. state}} & \multirow{2}{*}{\parbox{0.8cm}{Deg. state}} & \multirow{2}{*}{\parbox{0.9cm}{No. of nodes}} 
& \multicolumn{3}{c|}{DP} & \multicolumn{3}{c}{ECSCP} \\
\cline{5-10} 
&  &  &   & $R_{a}$  & $R_{b}$ & Energy & $R_{a}$  & $R_{b}$ & Energy   \\
\hline
\endfirsthead
\multicolumn{10}{r}%
{\tablename\ \thetable\ -- \textit{Continued from previous page}} \\
\hline
\multirow{2}{*}{\parbox{0.8cm}{Serial No.}} & \multirow{2}{*}{\parbox{0.8cm}{Ref. state}} & \multirow{2}{*}{\parbox{0.8cm}{Deg. state}} & \multirow{2}{*}{\parbox{0.9cm}{No. of nodes}} 
& \multicolumn{3}{c|}{DP} & \multicolumn{3}{c}{ECSCP} \\
\cline{5-10} 
&  &  &   & $R_{a}$  & $R_{b}$ & Energy & $R_{a}$  & $R_{b}$ & Energy   \\
\hline
\endhead
\hline \multicolumn{10}{r}{\textit{Continued on next page}} \\
\endfoot
\hline
\endlastfoot
$2as$ &  $2s$   &  $1s$ & 0 & 0              & 1.6982037372  & 0.1462201782   & 0              & 1.69818325927  & 0.14625383593 \\
$2bs$ &  $2s$   &  $1s$ & 0 & 1.6982037372   &  5            & 0.1462201782   & 1.69818325927  & 5              & 0.14625383593 \\
$2cs$ &  $2s$   &  $2s$ & 1 &  0             &  5            & 0.1462201782   & 0              & 5              & 0.14625383593 \\
$3as$ &  $3s$   &  $1s$ & 0 & 0              & 1.2580446107  & 1.0581877266   & 0              & 1.258034184915 & 1.05822024961 \\
$3bs$ &  $3s$   &  $1s$ & 0 & 1.2580446107   & 3.0543621890  & 1.0581877266   & 1.258034184915 & 3.054349992880 & 1.05822024961 \\
$3cs$ &  $3s$   &  $1s$ & 0 & 3.0543621890   & 5             & 1.0581877266   & 3.054349992880 &  5             & 1.05822024961 \\
$3ds$ &  $3s$   &  $2s$ & 1 & 0              & 3.0543621890  & 1.0581877266   & 0              & 3.054349992880 & 1.05822024961 \\
$3es$ &  $3s$   &  $2s$ & 1 & 1.2580446107   & 5             & 1.0581877266   & 1.258034184915 &  5             & 1.05822024961 \\
$3fs$ &  $3s$   &  $3s$ & 2 & 0              & 3.0543621890  & 1.0581877266   & 0              &  5             & 1.05822024961 \\
$4as$ &  $4s$   &  $1s$ & 0 & 0              & 0.99888647574 & 2.3872929368   & 0              & 0.998882989430 & 2.38732482829 \\
$4bs$ &  $4s$   &  $1s$ & 0 & 0.99888647574  & 2.28016476628 & 2.3872929368   & 0.998882989430 & 2.280158398620 & 2.38732482829 \\
$4cs$ &  $4s$   &  $1s$ & 0 & 2.28016476628  & 3.62627582410 & 2.3872929368   & 2.280158398620 & 3.626270386398 & 2.38732482829 \\
$4ds$ &  $4s$   &  $1s$ & 0 & 3.62627582410  & 5             & 2.3872929368   & 3.626270386398 & 5              & 2.38732482829 \\
$4es$ &  $4s$   &  $2s$ & 1 & 0              & 2.28016476628 & 2.3872929368   & 0              & 2.280158398620 & 2.38732482829 \\
$4fs$ &  $4s$   &  $2s$ & 1 & 0.99888647574  & 3.62627582410 & 2.3872929368   & 0.998882989430 & 3.626270386398 & 2.38732482829 \\
$4gs$ &  $4s$   &  $2s$ & 1 & 2.28016476628  & 5             & 2.3872929368   & 2.280158398620 &  5             & 2.38732482829 \\
$4hs$ &  $4s$   &  $3s$ & 2 &  0             & 3.62627582410 & 2.3872929368   & 0              & 3.626270386398 & 2.38732482829 \\
$4is$ &  $4s$   &  $3s$ & 2 & 0.99888647574  & 5             & 2.3872929368   & 0.998882989430 &  5             & 2.38732482829 \\
$4js$ &  $4s$   &  $4s$ & 3 & 0              & 5             & 2.3872929368   & 0              &  5             & 2.38732482829 \\
\hline
$3ap$ &  $3p$   &  $2p$ & 0 & 0              & 2.66019090630  & 0.712685104711 & 0              & 2.66017318418  & 0.7127180559   \\
$3bp$ &  $3p$   &  $2p$ & 0 & 2.66019090630  & 5              & 0.712685104711 & 2.66017318418  & 5              & 0.7127180559   \\
$3cp$ &  $3p$   &  $3p$ & 1 &  0             & 5              & 0.712685104711 & 0              & 5              & 0.7127180559   \\
$4ap$ &  $4p$   &  $2p$ & 0 &  0             & 1.887659760260 & 1.835390736624 & 0              & 1.887652169531 & 1.835423002175 \\
$4bp$ &  $4p$   &  $2p$ & 0 & 1.887659760260 & 3.43376514750  & 1.835390736624 & 1.887652169531 & 3.43375766692  & 1.835423002175 \\
$4cp$ &  $4p$   &  $2p$ & 0 & 3.43376514750  &   5            & 1.835390736624 & 3.43375766692  & 5              & 1.835423002175 \\
$4dp$ &  $4p$   &  $3p$ & 1 & 0              & 3.4337651475   & 1.835390736624 & 0              & 3.43375766692  & 1.835423002175 \\
$4ep$ &  $4p$   &  $3p$ & 1 & 1.887659760260 &  5             & 1.835390736624 & 1.887652169531 & 5              & 1.835423002175 \\
$4fp$ &  $4p$   &  $4p$ & 2 & 0              &  5             & 1.835390736624 & 0              & 5              & 1.835423002175 \\
$5ap$ &  $5p$   &  $2p$ & 0 & 0              & 1.475168797140 & 3.367619082432 & 0              & 1.475164887123 & 3.367650934911 \\
$5bp$ &  $5p$   &  $2p$ & 0 & 1.475168797140 & 2.644363878450 & 3.367619082432 & 1.475164887123 & 2.644358936263 & 3.367650934911 \\
$5cp$ &  $5p$   &  $2p$ & 0 & 2.644363878450 & 3.819016714020 & 3.367619082432 & 2.644358936263 & 3.819013041801 & 3.367650934911 \\
$5dp$ &  $5p$   &  $2p$ & 0 & 3.819016714020 &  5             & 3.367619082432 & 3.819013041801 & 5              & 3.367650934911 \\
$5ep$ &  $5p$   &  $3p$ & 1 & 0              & 2.644363878450 & 3.367619082432 & 0              & 2.644358936263 & 3.367650934911 \\
$5fp$ &  $5p$   &  $3p$ & 1 & 1.475168797140 & 3.819016714020 & 3.367619082432 & 1.475164887123 & 3.819013041801 & 3.367650934911 \\
$5gp$ &  $5p$   &  $3p$ & 1 & 2.644363878450 &    5           & 3.367619082432 & 2.644358936263 &   5            & 3.367650934911 \\
$5hp$ &  $5p$   &  $4p$ & 2 & 0              & 3.819016714020 & 3.367619082432 & 0              & 3.819013041801 & 3.367650934911 \\
$5ip$ &  $5p$   &  $4p$ & 2 & 1.475168797140 &   5            & 3.367619082432 & 1.475164887123 &  5             & 3.367650934911 \\
$5jp$ &  $5p$   &  $5p$ & 3 & 0      &      5                 & 3.367619082432 & 0              &  5             & 3.367650934911 \\
\hline
$4ad$ &  $4d$   &  $3d$ & 0 & 0              & 3.05208227850  & 1.244615818343 & 0              & 3.052071249402 & 1.244650639470 \\
$4bd$ &  $4d$   &  $3d$ & 0 & 3.05208227850  & 5              & 1.244615818343 & 3.052071249402 &   5            & 1.244650639470 \\
$4cd$ &  $4d$   &  $4d$ & 1 & 0              & 5              & 1.244615818343 &  0             &   5            & 1.244650639470 \\
$5ad$ &  $5d$   &  $3d$ & 0 & 0              & 2.24218003190  & 2.572144022112 &  0             & 2.242174038040 &  2.57217744067 \\
$5bd$ &  $5d$   &  $3d$ & 0 & 2.24218003190  & 3.63202185420  & 2.572144022112 & 2.242174038040 & 3.632016812010 &  2.57217744067 \\
$5cd$ &  $5d$   &  $3d$ & 0 & 3.63202185420  &   5            & 2.572144022112 & 3.63201681201  &   5            &  2.57217744067 \\
$5dd$ &  $5d$   &  $4d$ & 0 & 0              & 3.63202185420  & 2.572144022112 & 0              &  2.24217403804 &  2.57217744067 \\
$5ed$ &  $5d$   &  $4d$ & 0 & 2.24218003190  & 3.63202185420  & 2.572144022112 & 3.63201681201  &   5            &  2.57217744067 \\
$5fd$ &  $5d$   &  $5d$ & 0 & 0              & 5              & 2.572144022112 & 0              &   5            &  2.57217744067 \\
$6ad$ &  $6d$   &  $3d$ & 0 & 0              & 1.78281220402  & 4.306287086796 & 0              & 1.782814849720 &  4.306319753147 \\
$6bd$ &  $6d$   &  $3d$ & 0 & 1.78281220402  & 2.870891048410 & 4.306287086796 & 1.782814849720 & 2.870897278402 &  4.306319753147 \\
$6cd$ &  $6d$   &  $3d$ & 0 & 2.870891048410 & 3.937869825880 & 4.306287086796 & 2.870897278402 & 3.93788113360  &  4.306319753147 \\
$6dd$ &  $6d$   &  $3d$ & 0 & 3.937869825880 &  5             & 4.306287086796 & 3.93788113360  &  5             &  4.306319753147 \\
$6ed$ &  $6d$   &  $4d$ & 1 & 0              & 2.870891048410 & 4.306287086796 & 0              & 2.870897278402 &  4.306319753147 \\
$6fd$ &  $6d$   &  $4d$ & 1 & 1.78281220402  & 3.937869825880 & 4.306287086796 & 1.782814849720 & 3.93788113360  &  4.306319753147 \\
$6gd$ &  $6d$   &  $4d$ & 1 & 2.870891048410 &   5            & 4.306287086796 & 2.870897278402 &  5             &  4.306319753147 \\
$6hd$ &  $6d$   &  $5d$ & 2 & 0              & 3.937869825880 & 4.306287086796 & 0              & 3.93788113360  &  4.306319753147 \\
$6id$ &  $6d$   &  $5d$ & 2 & 1.78281220402  &  5             & 4.306287086796 & 1.782814849720 &  5             &  4.306319753147 \\
$6kd$ &  $6d$   &  $6d$ & 3 & 0              &   5            & 4.306287086796 & 0              &  5             &  4.306319753147 \\
\end{longtable}

In both DP and ECSCP number of bound states reduces with decrease in $\lambda$ values. Thus, it is more pertinent to examine the \emph{incidental degeneracy}
in plasma systems. Table~II, renders the incidental degeneracy for \emph{free} DP and ECSCP, taking (a) $n=2-4$ in $s$ states ($\ell=0$), (b) $n=3-5$ in $p$ 
states ($\ell=1$), and (c) $n=4-6$ in $d$ states ($\ell=2$), successively. Just to remind that, $R_{a}, R_{b}$ are chosen to be the nodal points of respective 
$s, p, d$ states in \emph{free} plasmas. In DP, at $n=2, \ell=0$ (energy$=-0.12007414$ a.u), a three fold degeneracy exists with one confined ($2as$), one 
left-confined ($2bs$) and one free ($2cs$) DP. Further, (a) $n=3, \ell=1$ (energy=$-0.05070822$ a.u.) and (b) $n=4, \ell=2$ (energy$=-0.0265028061$) states are three 
fold degenerate with one CS ($3ap$ or $4ad$), one LCS ($3bp$ or $4bd$) and one FC ($3cp$ or $4cd$) respectively. This degeneracy in \emph{shell-confined} DP involving 
$s, p, d$ states, however arises at $n=3,4,5$ consecutively with energies $\epsilon_{3s}=-0.0572018$ a.u., $\epsilon_{4p}=-0.02652592$ a.u. and $\epsilon_{5d}=-0.015406227$ a.u.         
In each of these three cases, there survives six degenerate states in DP, namely, confined ($3as,3ds$ or $4ap,4dp$ or $5ad,5dd$), shell-confined ($3bs$ or $4bp$ or $5bd$),
left-confined ($3cs,3es$ or $4cp,4ep$ or $5cd,5ed$) and free ($3fs$ or $4fp$ or $5fd$) DP. The incidental degeneracy in a \emph{shell-confined} 
$\ell$-state in DP appears at $n=(\ell+3)$. Further, for each of the following $4s$ ($\epsilon_{4s}=-0.02653747$ a.u.) or $5p$ ($\epsilon_{5p}=-0.015428340$ a.u.) or $6d$ 
($\epsilon_{6d}=-0.00947425$ a.u.) states there are ten degenerate states belonging to confined ($4as,4es,4hs$ or $5ap,5ep,5hp$ or $6ad,6ed,6hd$), shell-confined 
($4bs,4cs,4fs$ or $5bp,5cp,5fp$ or $6bd,6cd,6fd$), left-confined ($4ds,4gs,4is$ or $5dp,5gp,5ip$ or $6dd,6gd,6id$) and free ($4js$ or $5jp$ or $6jd$) DP respectively.     
Now moving to ECSCP system, one experiences exactly identical degeneracy pattern as DP, with explicit energy difference between the two. Barring an example we can mention that, 
akin to DP, in ECSCP such degeneracy involving a \emph{shell-confined} $\ell$-state arises at $n=(\ell+3)$. Therefore, one can aptly conjecture that, irrespective of the plasmas, 
incidental degeneracy for a given \emph{shell-confined} $\ell$-state arrives at $n=(\ell+3)$.        
     
\squeezetable          
\begin{longtable}{p{0.8cm}|p{0.8cm}|p{0.8cm}p{1.1cm}|ccc|ccc}
\caption{Incidental degeneracy in \emph{confined} DP and ECSCP under \emph{external electric field}, for $\lambda_{1},\lambda_2=200.0$ a.u. and $F$=0.1 a.u.} \\
\hline
\multirow{2}{*}{\parbox{0.8cm}{Serial No.}} & \multirow{2}{*}{\parbox{0.8cm}{Ref. state}} & \multirow{2}{*}{\parbox{0.8cm}{Deg. state}} & \multirow{2}{*}{\parbox{0.9cm}{No. of nodes}} 
& \multicolumn{3}{c|}{DP} & \multicolumn{3}{c}{ECSCP} \\
\cline{5-10} 
&  &  &   & $R_{a}$  & $R_{b}$ & Energy & $R_{a}$  & $R_{b}$ & Energy   \\
\hline
\endfirsthead
\multicolumn{10}{r}%
{\tablename\ \thetable\ -- \textit{Continued from previous page}} \\
\hline
\multirow{2}{*}{\parbox{0.8cm}{Serial No.}} & \multirow{2}{*}{\parbox{0.8cm}{Ref. state}} & \multirow{2}{*}{\parbox{0.8cm}{Deg. state}} & \multirow{2}{*}{\parbox{0.9cm}{No. of nodes}} 
& \multicolumn{3}{c|}{DP} & \multicolumn{3}{c}{ECSCP} \\
\cline{5-10} 
&  &  &   & $R_{a}$  & $R_{b}$ & Energy & $R_{a}$  & $R_{b}$ & Energy   \\
\hline
\endhead
\hline \multicolumn{10}{r}{\textit{Continued on next page}} \\
\endfoot
\hline
\endlastfoot
$2as$ &  $2s$   &  $1s$ & 0 & 0              & 1.63122916140  & 0.30421477724  & 0               & 1.63120322910  & 0.29925846975  \\
$2bs$ &  $2s$   &  $1s$ & 0 & 1.63122916140  &  120           & 0.30421477724  & 1.63120322910   &  125           & 0.29925846975  \\
$2cs$ &  $2s$   &  $2s$ & 1 & 0              &  $\infty$      & 0.30421477724  &  0              & $\infty$       & 0.29925846975  \\
$3as$ &  $3s$   &  $1s$ & 0 & 0              & 1.430987056806 & 0.64608902466  &  0              & 1.430960046710 & 0.64115528522  \\
$3bs$ &  $3s$   &  $1s$ & 0 & 1.430987056806 & 3.98416222870  & 0.64608902466  & 1.430960046710  & 3.98405258470  & 0.64115528522  \\
$3cs$ &  $3s$   &  $1s$ & 0 & 3.98416222870  &  150           & 0.64608902466  & 3.98405258470   &   125          & 0.64115528522  \\
$3ds$ &  $3s$   &  $2s$ & 1 & 0              & 3.98416222870  & 0.64608902466  &  0              & 3.98405258470  & 0.64115528522  \\
$3es$ &  $3s$   &  $2s$ & 1 & 1.430987056806 & 150            & 0.64608902466  & 1.430960046710  &   150          & 0.64115528522  \\
$3fs$ &  $3s$   &  $3s$ & 2 & 0              & $\infty$       & 0.64608902466  &  0              &  $\infty$       & 0.64115528522  \\
$4as$ &  $4s$   &  $1s$ & 0 & 0              & 1.323704568110 & 0.91194766539  &  0              & 1.323677129230 & 0.90703311507  \\
$4bs$ &  $4s$   &  $1s$ & 0 & 1.323704568110 & 3.437749731312 & 0.91194766539  & 1.323677129230  & 3.43765212650  & 0.90703311507  \\
$4cs$ &  $4s$   &  $1s$ & 0 & 3.437749731312 & 6.2231958170   & 0.91194766539  & 3.43765212650   & 6.22299514840  & 0.90703311507  \\
$4ds$ &  $4s$   &  $1s$ & 0 & 6.2231958170   & 150            & 0.91194766539  & 6.22299514840   &  125           & 0.90703311507  \\
$4es$ &  $4s$   &  $2s$ & 1 & 0              & 3.437749731312 & 0.91194766539  &  0              & 3.43765212650  & 0.90703311507  \\
$4fs$ &  $4s$   &  $2s$ & 1 & 1.323704568110 & 6.2231958170   & 0.91194766539  & 1.323677129230  & 6.22299514840  & 0.90703311507  \\
$4gs$ &  $4s$   &  $2s$ & 1 & 3.437749731312 &  150           & 0.91194766539  & 3.43765212650   &  150           & 0.90703311507  \\
$4hs$ &  $4s$   &  $3s$ & 2 & 0              & 6.2231958170   & 0.91194766539  &  0              & 6.22299514840  & 0.90703311507  \\
$4is$ &  $4s$   &  $3s$ & 2 & 1.323704568110 &  180           & 0.91194766539  &  1.323677129230 &   150          & 0.90703311507  \\
$4js$ &  $4s$   &  $4s$ & 3 & 0              &  $\infty$      & 0.91194766539  &   0             &  $\infty$       & 0.90703311507  \\
\hline
$3ap$ &  $3p$   &  $2p$ & 0 & 0                & 3.1735238880  & 0.57546550714 & 0               &  3.17343008710 & 0.57052461177 \\
$3bp$ &  $3p$   &  $2p$ & 0 & 3.1735238880     &  150          & 0.57546550714 & 3.1734300871    &     125        & 0.57052461177 \\
$3cp$ &  $3p$   &  $3p$ & 1 & 0                &  $\infty$     & 0.57546550714 &  0              &  $\infty$      & 0.57052461177 \\
$4ap$ &  $4p$   &  $2p$ & 0 & 0                & 2.6815598797  & 0.84563053412 &  0              & 2.68147834530  & 0.84070941159 \\
$4bp$ &  $4p$   &  $2p$ & 0 & 2.6815598797     & 5.5042196423  & 0.84563053412 & 2.68147834530   &  5.50403433220 & 0.84070941159 \\
$4cp$ &  $4p$   &  $2p$ & 0 & 5.5042196423     & 180           & 0.84563053412 & 5.50403433220   &   125          & 0.84070941159 \\
$4dp$ &  $4p$   &  $3p$ & 1 & 0                &  5.5042196423 & 0.84563053412 &  0              & 5.50403433220  & 0.84070941159 \\
$4ep$ &  $4p$   &  $3p$ & 1 & 2.6815598797     &  180          & 0.84563053412 &  2.68147834530  &   150          & 0.84070941159 \\
$4fp$ &  $4p$   &  $4p$ & 2 & 0                & $\infty$      & 0.84563053412 &  0              &  $\infty$      & 0.84070941159 \\
$5ap$ &  $5p$   &  $2p$ & 0 & 0                & 2.41957855790 & 1.0779685915  &  0              & 2.419503441880 & 1.0730649761 \\
$5bp$ &  $5p$   &  $2p$ & 0 & 2.41957855790    & 4.73787060730 & 1.0779685915  & 2.419503441880  & 4.737710002070 & 1.0730649761 \\
$5cp$ &  $5p$   &  $2p$ & 0 & 4.73787060730    & 7.6216242920  & 1.0779685915  & 4.737710002070  & 7.6213531670   & 1.0730649761 \\
$5dp$ &  $5p$   &  $2p$ & 0 & 7.6216242920     &   200         & 1.0779685915  & 7.6213531670    &  150           & 1.0730649761 \\
$5ep$ &  $5p$   &  $3p$ & 1 & 0                & 4.73787060730 & 1.0779685915  &  0              & 4.737710002070 & 1.0730649761 \\
$5fp$ &  $5p$   &  $3p$ & 1 & 2.41957855790    & 7.6216242920  & 1.0779685915  & 2.419503441880  & 7.6213531670   & 1.0730649761 \\
$5gp$ &  $5p$   &  $3p$ & 1 & 4.73787060730    &  200          & 1.0779685915  & 4.737710002070  &  160           & 1.0730649761 \\
$5hp$ &  $5p$   &  $4p$ & 2 & 0                & 7.6216242920  & 1.0779685915  &   0             & 7.6213531670   & 1.0730649761 \\
$5ip$ &  $5p$   &  $4p$ & 2 & 2.41957855790    &  200          & 1.0779685915  &  2.419503441880 &   180          & 1.0730649761 \\
$5jp$ &  $5p$   &  $5p$ & 3 &  0               &  $\infty$     & 1.0779685915  &   0             &  $\infty$      & 1.0730649761 \\
\hline
$4ad$ &  $4d$   &  $3d$ & 0 & 0                & 4.37176436570 & 0.76153364255 & 0               & 4.371614629150 & 0.7566050503 \\
$4bd$ &  $4d$   &  $3d$ & 0 & 4.37176436570    & 200           & 0.76153364255 & 4.371614629150  &  130           & 0.7566050503 \\
$4cd$ &  $4d$   &  $4d$ & 1 &  0               & $\infty$      & 0.76153364255 &   0             &   $\infty$     & 0.7566050503 \\
$5ad$ &  $5d$   &  $3d$ & 0 &  0               &  3.6994249718 & 0.99951100719 &   0             & 3.6992974680   & 0.99460049179 \\
$5bd$ &  $5d$   &  $3d$ & 0 &  3.6994249718    & 6.676090403   & 0.99951100719 & 3.6992974680    & 6.675849412170 & 0.99460049179 \\
$5cd$ &  $5d$   &  $3d$ & 0 &  6.676090403     &   200         & 0.99951100719 &  6.675849412170 &     180        & 0.99460049179 \\
$5dd$ &  $5d$   &  $4d$ & 0 &  0               &  6.676090403  & 0.99951100719 &   0             & 6.675849412170 & 0.99460049179 \\
$5ed$ &  $5d$   &  $4d$ & 0 & 3.6994249718     &   200         & 0.99951100719 & 3.6992974680    &       200      & 0.99460049179 \\
$5fd$ &  $5d$   &  $5d$ & 0 &  0               &  $\infty$     & 0.99951100719 &   0             & $\infty$       & 0.99460049179 \\
$6ad$ &  $6d$   &  $3d$ & 0 &  0               &  3.3301902855 & 1.21275230925 &  0              &  3.3300747810  & 1.2078581777   \\
$6bd$ &  $6d$   &  $3d$ & 0 &  3.33019028550   & 5.77461618903 & 1.21275230925 &  3.3300747810   &  5.7744075560  & 1.2078581777   \\
$6cd$ &  $6d$   &  $3d$ & 0 &  5.774616189030  & 8.72983376980 & 1.21275230925 &  5.7744075560   & 8.7295104230   & 1.2078581777   \\
$6dd$ &  $6d$   &  $3d$ & 0 &  8.72983376980   &  200          & 1.21275230925 & 8.7295104230    &  180           & 1.2078581777   \\
$6ed$ &  $6d$   &  $4d$ & 1 &  0               & 5.77461618903 & 1.21275230925 &  0              &  5.7744075560  & 1.2078581777   \\
$6fd$ &  $6d$   &  $4d$ & 1 & 3.33019028550    & 8.72983376980 & 1.21275230925 &  3.3300747810   &  8.7295104230  & 1.2078581777   \\
$6gd$ &  $6d$   &  $4d$ & 1 & 5.774616189030   &  200          & 1.21275230925 &  5.7744075560   &  180           & 1.2078581777   \\
$6hd$ &  $6d$   &  $5d$ & 2 & 0                & 8.72983376980 & 1.21275230925 &  0              & 8.7295104230   & 1.2078581777   \\
$6id$ &  $6d$   &  $5d$ & 2 & 3.33019028550    &  200          & 1.21275230925 &  3.3300747810   &  200           & 1.2078581777   \\
$6kd$ &  $6d$   &  $6d$ & 3 &   0              &  $\infty$     & 1.21275230925 &  0              &  $\infty$      & 1.2078581777   \\
\end{longtable}

At this point, we aim to verify the impact of high pressure and external electric field on such degeneracy pattern in DP and ECSCP.  
Table~III, imprints the incidental degeneracy for \emph{confined} DP and ECSCP at $r_{c}=5$ a.u., for $2s, 3s, 4s, 3p, 4p, 5p, 4d, 5d, 6d$ states. 
As usual, $R_{a}, R_{b}$ are fixed at the nodal points of successive $s, p, d$ orbitals in \emph{confined} plasmas. Similarly, Table~IV displays the 
same for DP and ECSCP under the impression of external electric field ($F \cos \theta =0.1$ a.u.). Detailed analysis reveals that, in high 
pressure condition and in external electric field one encounters exactly uniform pattern of degeneracy as their respective \emph{free} counterpart, 
with obvious change in energy between them. Therefore, incidental degeneracy remains unaffected with these two external effects. This again promptly 
asserts the implicitness of this degeneracy. 

Now, analyzing Tables~II-IV, it can be extracted that, there arises $\frac{(n-\ell)(n-\ell+1)}{2}$ number of degenerate states 
in GCS. Therefore, for a certain $n$-state of DP or ECSCP, the total count of GCS states becomes, $\frac{(n-\ell)(n-\ell+1)}{2}$. It suggests that, 
this number depends on both $n, \ell$ quantum numbers. Now, we can, calculate the contribution of each of these four systems in this  degeneracy, as 
given below. 

\begin{enumerate}[label=(\Roman*)]
\item 
\textbf{FC:} At a given $(n,\ell)$-state, there exists only one energy states. 
\item
\textbf{CS:} An $(n,\ell)$-orbital contributes $(n-\ell-1)$ number of degenerate states.
\item
\textbf{SCC:} $(n,\ell)$-state contributes as $\frac{(n-\ell-2)(n-\ell-1)}{2}$.     
\item
\textbf{LCS:} A particular $(n,\ell)$ orbital contributes $(n-\ell-1)$ degenerate states.
\end{enumerate}  
  
In confined plasmas, LCS modifies to SCS with $R_{b}$ is kept fixed at a certain value. Thus, SCC contributes, 
\begin{equation}
\begin{aligned}
& \frac{(n-\ell-1)(n-\ell-2)}{2}+(n-\ell-1) \\
& = (n-\ell-1)\left[1+\frac{(n-\ell-2)}{2}\right] \\
& = \frac{(n-\ell)(n-\ell-1)}{2}.  
\end{aligned}
\end{equation}

On the basis of above discussion and Tables~II-IV, one can trace out certain characteristics. In order to simplify, we mention $n, \ell$ to refer principal 
and orbital quantum number of \emph{free} plasmas, whereas $n_{k}, \ell_{j}$ indicates the same for other three systems ($k, j$ are integers). 

\begin{enumerate}[label=(\roman*)]
\item
Each $\ell$-state having fixed $n$ value, contributes $\frac{(n-\ell)(n-\ell+1)}{2}$ number of GCS states. 
\item
At a certain $\ell$ the count of such degenerate state increases with $n$. On the contrary, at a certain $n$, the number of degenerate states reduce with rise in 
$\ell$.
\item 
In SCC degeneracy appears at $n=3$.  
\item
Two arbitrary states $(n_{1},\ell_{1})$, $(n_{2},\ell_{2})$ become degenerate when $n_{1} < n, \ell_{1} < n$ and 
$n_{2} < n, \ell_{2} < n$. They may be associated to any of the systems in GCS, except FC. 
\end{enumerate}
   
\begingroup                    
\begin{table}
\caption{Additional degeneracy in \emph{confined} DP and ECSCP, for $\lambda_{1},\lambda_2=200$ a.u.}
\centering
\begin{tabular}{ll|lll|lll}
\hline	
& &\multicolumn{3}{c|}{DP} & \multicolumn{3}{c}{ECSCP} \\
\hline
Serial No. & State & $R_{a}$  & $R_{b}$ & Energy & $R_{a}$  & $R_{b}$ & Energy  \\
\hline
$1$       &  $1s$ & 0  & 1            & 2.3789850177 &  0       & 1            & 2.3789908559 \\
$2$       &  $2s$ & 0  & 2.2832202233 & 2.3789850177 &  0       & 2.2832234381 & 2.3789908559 \\
$3$       &  $2p$ & 0  & 1.7024233032 & 2.3789850177 &  0       & 1.7024251301 & 2.3789908559 \\
$4$       &  $3s$ & 0  & 3.6315075428 & 2.3789850177 &  0       & 3.6315185462 & 2.3789908559 \\
$5$       &  $3p$ & 0  & 3.0758583379 & 2.3789850177 &  0       & 3.0758664112 & 2.3789908559 \\
$6$       &  $3d$ & 0  & 2.3198593110 & 2.3789850177 &  0       & 2.3198645970 & 2.3789908559 \\
$7$       &  $4s$ & 0  & 5.0075095717 & 2.3789850177 &  0       & 5.0075332221 & 2.3789908559 \\
$8$       &  $4p$ & 0  & 4.4617689881 & 2.3789850177 &  0       & 4.4617880726 & 2.3789908559 \\
$9$       &  $4d$ & 0  & 3.7617324892 & 2.3789850177 &  0       & 3.7617470592 & 2.3789908559 \\
$10$      &  $4f$ & 0  & 2.9013975139 & 2.3789850177 &  0       & 2.9014076831 & 2.3789908559 \\
..        & ..    & .. & ..             & ..             & ..       & ..      & ..             \\
..        & ..    & .. & ..             & ..             & ..       & ..      & ..             \\  
\hline
$1$       &  $1s$ & 0  & 2              & $-$0.1200107236 & 0     & 2                & $-$0.12000003625  \\
$2$       &  $2s$ & 0  & 17.9007405190 & $-$0.1200107236 & 0     & 22.5883720000 & $-$0.12000003625  \\ 
$3$       &  $2p$ & 0  & 17.0684450000 & $-$0.1200107236 & 0     & 21.8984020000 & $-$0.12000003625  \\
\hline
$1$       &  $1s$ & 0  & 3             & $-$0.4189816690 & 0     & 3       & $-$0.418967355112  \\
\hline
\end{tabular}
\end{table}  
\endgroup  

Apart from the incidental degeneracy, here we point out the existence of another degeneracy in confined plasmas ($R_{a}=0$).
Table~V, exhibit such degeneracy pattern in confined DP and ECSCP involving $1s$ state. At $R_{b}=1$, the energy ($\epsilon_{1s}=2.3789850177$) 
of $1s$-state becomes degenerate with $2s$ ($R_{b}=2.2832202233$), $2p$ ($R_{b}=1.7024233032$), $3s$ ($R_{b}=3.6315075428$), ...., $4f$ 
($R_{b}=2.9013975139$), and so on. At, this $R_{a}$ value, $\epsilon_{1s}$ is positive and greater than $\epsilon^{free}_{n,\ell}$. Here, 
$\epsilon^{free}_{n,\ell}$ represent the energy of the $(n,\ell)$ state in absence
of pressure.  However, at $R_{b}=2$, 1s-state ($\epsilon_{1s}=-0.1200107236$) is degenerate with only
$2s$ ($R_{b}=17.9007405190$) and $2p$ ($R_{b}=17.0684450000$) states. It happens because at $R_{b}=2$, $\epsilon_{1s}$ is only higher than the 
energy of $\epsilon^{free}_{2s}, \epsilon^{free}_{2p}$. Finally, at $R_{b}=3$, $\epsilon_{1s}$ becomes non-degenerate. Similar observation can also be obtain 
for excited states. In essence it can be stated that, when $\epsilon_{1s} > 0$, then there exists infinite number of such degenerate states. However, when 
$\epsilon_{1s} < 0$ region, with decrease in energy 
value count of such degenerate state declines. After a certain limiting value, it becomes non degenerate. This type of degeneracy occurs only in confined 
systems and the count of such degenerate state complete depends on the strength of confinement. With increase in pressure, the energy of $1s$-state increases, 
as consequence number of degenerate state increases. More importantly, here we have presented the results for $1s$-state. However, similar observation can also be 
recorded for an arbitrary bound state.

\begingroup              
\squeezetable
\begin{table}
\caption{$f^{(1)}$ involving $1s,2s,2p$ states in \emph{confined} DP and ECSCP for $\lambda_{1},\lambda_2=200$ a.u.}
\centering
\begin{ruledtabular}
\begin{tabular}{ll|ll|ll|llll}
\multicolumn{10}{c}{DP} \\
\hline
$R_{a}$  & $R_{b}$ & $1s \rightarrow 2p$  & $1s \rightarrow 3p$ & $2s \rightarrow 2p$ & $2s \rightarrow 3p$ & 
$2p \rightarrow 1s$  & $2p \rightarrow 2s$ & $2p \rightarrow 3d$ & $2p \rightarrow 4d$ \\
\hline	
0    & 1  & 0.9845583736  & 0.007725936 & $-$0.6082578489 & 1.5603265067 & $-$0.3281861245 & 0.2027526163 & 1.084824819 & 0.018576821 \\
0.1  & 1  & 0.8991021968  & 0.091177999 & $-$0.5487515200 & 1.2675987342 & $-$0.2997007322 & 0.1829171733 & 1.078261447 & 0.025793358 \\
0.2  & 1  & 0.8115882841  & 0.176644625 & $-$0.4643464211 & 1.0121522214 & $-$0.2705294280 & 0.1547821403 & 1.044480208 & 0.059665545 \\
0.5  & 1  & 0.6974611899  & 0.289644039 & $-$0.3508420476 & 0.7344790098 & $-$0.2324870633 & 0.1169473492 & 0.929550287 & 0.173160326 \\
0.8  & 1  & 0.6699145798  & 0.316999576 & $-$0.3234507070 & 0.6737046104 & $-$0.2233048599 & 0.1078169023 & 0.893218984 & 0.209183954  \\
\hline                                                                                      
0    & 5  & 0.8488024520  & 0.108272180 & $-$0.4563748356 & 1.4233367326 & $-$0.2829341506 & 0.1521249452 & 1.103032718 & 0.001023063 \\
0.5  & 5  & 0.93301292953 & 0.053537177 & $-$0.5865015753 & 1.3312315839 & $-$0.3110043098 & 0.1955005251 & 1.097736184 & 0.008041007 \\
1    & 5  & 0.82132893350 & 0.164523624 & $-$0.4739305826 & 1.0274313977 & $-$0.2737763111 & 0.1579768608 & 1.106595561 & 0.000413399 \\
2    & 5  & 0.72015045177 & 0.266781857 & $-$0.3732478855 & 0.7855584038 & $-$0.2400501505 & 0.1244159618 & 0.958259417 & 0.144671064 \\
3    & 5  & 0.68357364638 & 0.303403595 & $-$0.3370006030 & 0.7035722963 & $-$0.2278578821 & 0.1123335343 & 0.911368314 & 0.191162267 \\
4.5  & 5  & 0.66739177323 & 0.319504855 & $-$0.3209450048 & 0.6682356487 & $-$0.2224639244 & 0.1069816682 & 0.889855689 & 0.212523233 \\
\hline                                                      
0   & 10 & 0.49197804318  & 0.258217079 & $-$0.0777278384 & 0.9674872235 & $-$0.1639926810 & 0.0259092794 & 1.070082794 & 0.027997652 \\
0.5 & 10 & 0.97045011671  & 0.014376086 & $-$0.6209901235 & 1.5633135494 & $-$0.3234833722 & 0.2069967078 & 1.086251699 & 0.016845793 \\
1   & 10 & 0.94768087310  & 0.024025469 & $-$0.5899418568 & 1.3674404123 & $-$0.3158936243 & 0.1966472856 & 1.106595561 & 0.000413399 \\
3   & 10 & 0.75955594681  & 0.224597411 & $-$0.4104142965 & 0.8749891665 & $-$0.2531853156 & 0.1368047655 & 1.004384636 & 0.098397827 \\
5   & 10 & 0.69774049664  & 0.289034898 & $-$0.3508574954 & 0.7348647740 & $-$0.2325801655 & 0.1169524984 & 0.929957661 & 0.172609822 \\
7.5 & 10 & 0.67206042643  & 0.314861162 & $-$0.3255754666 & 0.6783672042 & $-$0.2240201421 & 0.1085251555 & 0.896078528 & 0.206340948 \\
9.5 & 10 & 0.66683857872  & 0.320054255 & $-$0.3203956825 & 0.6670385359 & $-$0.2222795262 & 0.1067985608 & 0.889118104 & 0.213255697 \\
\hline
 \multicolumn{10}{c}{ECSCP} \\
\hline
$R_{a}$  & $R_{b}$ & $1s \rightarrow 2p$  & $1s \rightarrow 3p$ & $2s \rightarrow 2p$ & $2s \rightarrow 3p$ & 
$2p \rightarrow 1s$  & $2p \rightarrow 2s$ & $2p \rightarrow 3d$ & $2p \rightarrow 4d$ \\
\hline
0    & 1  & 0.9845583944 & 0.0077259195 & $-$0.6082578857 & 1.5603265595 & $-$0.3281861314 & 0.2027526285 & 1.084824835 & 0.018576809 \\
0.1  & 1  & 0.8991022160 & 0.0911779811 & $-$0.5487515470 & 1.2675987702 & $-$0.2997007386 & 0.1829171823 & 1.078261461 & 0.025793346 \\
0.2  & 1  & 0.8115882910 & 0.1766446184 & $-$0.4643464305 & 1.0121522331 & $-$0.2705294303 & 0.1547821435 & 1.044480217 & 0.059665537 \\
0.5  & 1  & 0.6974611901 & 0.2896440393 & $-$0.3508420478 & 0.7344790100 & $-$0.2324870633 & 0.1169473492 & 0.929550287 & 0.173160326 \\
0.8  & 1  & 0.6699145791 & 0.3169995763 & $-$0.3234507070 & 0.6737046103 & $-$0.2233048597 & 0.1078169023 & 0.893218984 & 0.209183954 \\
\hline                                                   
0    & 5  & 0.8487992934 & 0.1082749684 & $-$0.4563746338 & 1.4233366775 & $-$0.2829330978 & 0.1521248779 & 1.103033440 & 0.001022600 \\
0.5  & 5  & 0.9330138952 & 0.0535357427 & $-$0.5865027973 & 1.3312341111 & $-$0.3110046317 & 0.1955009324 & 1.097737370 & 0.008040088 \\
1    & 5  & 0.7201505110 & 0.2667817486 & $-$0.4739309943 & 1.0274323378 & $-$0.2737764770 & 0.1579769981 & 1.055607457 & 0.049069981 \\
2    & 5  & 0.6835736500 & 0.3034035886 & $-$0.3732479200 & 0.7855585007 & $-$0.2400501703 & 0.1244159733 & 0.958259517 & 0.144670958 \\
3    & 5  & 0.6835736500 & 0.3034035886 & $-$0.3370006045 & 0.7035723022 & $-$0.2278578833 & 0.1123335348 & 0.911368319 & 0.191162261 \\
4.5  & 5  & 0.6673917703 & 0.3195048550 & $-$0.3209450048 & 0.6682356472 & $-$0.2224639234 & 0.1069816682 & 0.889855688 & 0.212523233 \\
\hline                                               
0   & 10 & 0.492037585 & 0.2581751997 & $-$0.0778155020 & 0.9675474065 & $-$0.1640125285 & 0.0259385006 & 1.070064435 & 0.028015562 \\
0.5 & 10 & 0.970452919 & 0.0143709867 & $-$0.6209863259 & 1.5633183645 & $-$0.3234843064 & 0.2069954419 & 1.086240012 & 0.016858413 \\
1   & 10 & 0.947674297 & 0.0240247854 & $-$0.5899250652 & 1.3674340783 & $-$0.3158914323 & 0.1966416884 & 1.106595224 & 0.000414800 \\
3   & 10 & 0.759556484 & 0.2245945414 & $-$0.4104128653 & 0.8749903496 & $-$0.2531854949 & 0.1368042884 & 1.004386314 & 0.098395452 \\
5   & 10 & 0.697740596 & 0.2890345618 & $-$0.3508574039 & 0.7348649380 & $-$0.2325801989 & 0.1169524679 & 0.929957822 & 0.172609559 \\
7.5 & 10 & 0.672060427 & 0.3148611579 & $-$0.3255754655 & 0.6783672047 & $-$0.2240201425 & 0.1085251551 & 0.896078529 & 0.206340945 \\
9.5 & 10 & 0.666838573 & 0.3200542553 & $-$0.3203956826 & 0.6670385259 & $-$0.2222795243 & 0.1067985608 & 0.889118103 & 0.213255697 \\
\end{tabular}
\end{ruledtabular}
\end{table}  
\endgroup          

\subsection{Multipole oscillator strength and polarizability}  
This section is divided into two parts. At first we report $f^{(1)}$ and $\alpha^{(1)}$ for confined DP and ECSCP, in some low-lying states. 
Finally, we discuss the impact of external electric field on $f^{(1)}$ and $\alpha^{(1)}$ for these two confined plasmas. Except 
$f^{(1)}, \alpha^{(1)}$ of $1s,2s$ in confined DP and ECSCP, no results are presented so far in any of the GCS plasmas \cite{mukherjee21}. 
Coincidentally, that calculations are performed by the present author using GPS method. However, the impact of external electric field on 
these $f^{(1)},\alpha^{(1)}$ are not been reported before. As a consequence, no such literature is available for comparison.

At the beginning, we point out that, the oscillator strength sum rule, Eq.~(\ref{eq:9}) is obeyed by all states in confined DP and 
ECSCP. $f^{(k)}$ measures the probability of transition from an initial to final state. It is the ratio involving the quantum 
mechanical transition rate and the classical absorption/emission rate connecting a single electron oscillator having same frequency of transition. It is
a dimensionless quantity. It becomes $-$ve for emission and $+$ve for absorption. Further, an emissive state with small value of oscillator strength 
explains that, non-radiative decay outpaces radiative decay \cite{robert62}.          

The selection rule for dipole transition is, $\Delta \ell=\pm 1$. As a consequence, from an $s$ state, transition is only possible to a $p$ state. 
But from $p$ state, transition may takes place to $\ell =0,2$ states. Table~VI demonstrates estimated $f^{(1)}$ for $1s,2s,2p$ involving 
$n,\ell \rightarrow n^{'},(\ell+1)$ ($n=1,2;n^{'}=2,3,4$) transitions involving DP and ECSCP. In either of the plasmas, SCC outcomes are presented for 
three $R_{b}$, namely, $1,5,10$; for each $R_{b}$, $R_{a}$ changes from $0-R_{b}$. Here, CS situation is achieved at $R_{a}=0$, consequently, $r_{c}=R_{b}=1,5,10$.
$f^{(1)}_{1s \rightarrow 2p}$ in both confined DP and ECSCP decreases with rise in $r_{c}$ to merge to respective free values. In 
\emph{shell-confined} condition, for $R_{b}=1$ it declines with progress in $R_{a}$. At $R_{b}=5,10$ it increases to reach a maximum then decreases. 
The behavior of $f^{(1)}_{1s \rightarrow 3p}$ is distinctly different from $f^{(1)}_{1s \rightarrow 2p}$. In confined DP and ECSCP completely 
opposite trend is observed$--$ it increases with advancement of $r_{c}$ to converge to their free limits. In SCC, at $R_{b}=1$, it advances with 
$R_{a}$. However, at $R_{b}=5,10$ it declines to reach a minimum then increases. $f^{(1)}_{2s \rightarrow 2p}$ is always $(-)$ve, which indicates that, 
except \emph{free} plasmas, the former has higher energy than latter. At $R_{a}=0$, its absolute value abates with progress in $r_{c}=R_{b}$. Interestingly 
in SCC, at $R_{b}=1$, its magnitude decreases with enhancement of $R_{a}$. Beside this, at $R_{b}=5,10$, its absolute value increases to climb a maximum then 
declines with growth in $R_{a}$. For $f^{(1)}_{2s \rightarrow 3p}$ transition, $f^{(1)}$ in \emph{confined} plasmas declines with progress in $r_{c}$ to convene 
to their free values. Further, at $R_{b}=1,5$, it again abates with rise in $R_{a}$. Conversely, at $R_{b}=10$, it attains a maximum then decreases. Similar to 
$f^{(1)}$ in $2s \rightarrow 2p$ transition, $f^{(1)}_{2p \rightarrow 1s}$ is always $-$ve. Moreover, in both \emph{confined} and \emph{shell-confined} plasmas
it portrays exactly identical pattern to $f^{(1)}_{2s \rightarrow 2p}$. Although they possesses different numerical values. In case of $2p \rightarrow 2s$ 
transition, $f^{(1)}$ decreases with rise in $r_{c}$. However, in SCC, it imprints similar behavior to $f^{(1)}_{1s \rightarrow 3p}$. Now, for 
$f^{(1)}_{2p \rightarrow 3d}$, in CS we observe a maximum. It is noteworthy to mention that, in SCC, it imprints similar pattern to $1s \rightarrow 3p$ transition. 
Importantly, in \emph{confined} plasmas $f^{(1)}_{2p \rightarrow 4d}$ behaves exactly reverse to $f^{(1)}_{2p \rightarrow 3d}$. Further, in \emph{shell-confined}
situation, at $R_{b}=1,5$, it advances with rise in $R_{a}$. On the contrary, at $R_{b}=10$, it falls off to a minimum then increases. 
                         
\begingroup     
\squeezetable
\begin{table}
\caption{$\alpha^{(1)}$ involving $1s,2s,3s,4s,2p,3d$ states for \emph{confined} DP and ECSCP for $\lambda_{1},\lambda_2=200$ a.u.}
\centering
\begin{ruledtabular}
\begin{tabular}{ll|l|llll|ll}
\multicolumn{9}{c}{DP} \\
\hline
$R_{a}$  & $R_{b}$ & $V$ & $\alpha^{(1)}_{1s}$ & $\alpha^{(1)}_{2s}$ & $\alpha^{(1)}_{3s}$ & $\alpha^{(1)}_{4s}$ & $\alpha^{(1)}_{2p}$ &
$\alpha^{(1)}_{3d}$  \\
\hline
0    & 1 & 1        & 0.02879203   & 0.00441401     & 0.00188747    & 0.00105362           & 0.01719826  & 0.00894345     \\
0.1  & 1 & 0.999    & 0.04759423   & 0.02720296     & 0.02357305    & 0.02219161           & 0.01004333  & 0.00815930     \\
0.2  & 1 & 0.992    & 0.07284697   & 0.05487310     & 0.05124971    & 0.04988689           & 0.00583027  & 0.00565907     \\
0.5  & 1 & 0.875    & 0.20188347   & 0.19133640     & 0.18923265    & 0.18848056           & 0.00083203  & 0.00083479     \\
0.8  & 1 & 0.488    & 0.43528355   & 0.43282925     & 0.43236927    & 0.43220785           & 0.00002108  & 0.00002108     \\
0.9  & 1 & 0.271    & 0.54241441   & 0.54172916     & 0.54160189    & 0.54155732           & 0.00000131  & 0.00000131     \\
\hline
0    & 5 & 125      &   3.42266613 & $-$21.10698996 & $-$5.69158760 & $-$2.65406011        & 19.323064623  & 7.211907222       \\
0.5  & 5 & 124      &  19.24630056 &    16.11989657 &   15.00888999 &   14.25741078        &  6.325506169  & 6.027928662       \\
1    & 5 & 117      &  38.31053635 &    34.92820757 &   32.75176993 &   31.71721197        &  3.578607933  & 3.72435036       \\
2    & 5 & 109.375  &  90.38405129 &    85.34533557 &   83.48821755 &   82.73470233        &  1.08354091  &  1.09660350      \\
3    & 5 & 98       & 165.90913640 &   161.85818857 &  160.86288662 &  160.49302864        &  0.21171103  &  0.21200761      \\
4.5  & 5 & 33.875   & 339.00622748 &   338.58155053 &  338.50167645 &  338.47363010        &  0.00082350  &  0.00082314      \\
\hline
0    & 10 & 1000    &    4.49736499 & $-$2089.73701909 & $-$376.88040323 & $-$143.24784214 & 794.443870770 & 171.83606073      \\
0.5  & 10 & 999.875 &   57.87595003 &      87.13375737 &    107.30929287 &    110.53658538 & 107.422952610 & 152.90582154      \\
1    & 10 & 999     &  163.68450247 &     277.64800144 &    254.02359344 &    238.62785299 &  80.715953070 & 113.77076642      \\
3    & 10 & 973     &  900.96379839 &     936.69782121 &    895.51223588 &    876.02414752 &  31.490416258 &  33.27278240      \\
5    & 10 & 875     & 1969.38863563 &    1925.74517368 &   1900.37617832 &   1889.89088703 &   8.272785518 &   8.32198498      \\
7.5  & 10 & 657     & 3873.03746168 &    3841.48716664 &   3834.24959362 &   3831.60721504 &   0.51510315 &    0.515184429     \\
9.5  & 10 & 142.625 & 6023.02648886 &    6021.22915052 &   6020.89380551 &   6020.77623107 &   0.00082430 &    0.00083429     \\
\hline
\multicolumn{9}{c}{ECSCP} \\
\hline
$R_{a}$  & $R_{b}$ & $V$ & $\alpha^{(1)}_{1s}$ & $\alpha^{(1)}_{2s}$ & $\alpha^{(1)}_{3s}$ & $\alpha^{(1)}_{4s}$ & $\alpha^{(1)}_{2p}$ &
$\alpha^{(1)}_{3d}$  \\
\hline
0    & 1 & 1        & 0.02879202  & 0.00441401 & 0.00188746 &  0.0010536  & 0.01719827 & 0.00894345 \\
0.1  & 1 & 0.999    & 0.04759422  & 0.02720296 & 0.02357305 &  0.0221916  & 0.01004333 & 0.00815930 \\
0.2  & 1 & 0.992    & 0.07284696  & 0.05487311 & 0.05124971 &  0.0498868  & 0.00583027 & 0.00565907 \\
0.5  & 1 & 0.875    & 0.20188347  & 0.19133640 & 0.18923265 &  0.1884805  & 0.00083203 & 0.00083479 \\
0.8  & 1 & 0.488    & 0.43528355  & 0.43282925 & 0.43236927 &  0.432207   & 0.00002108 & 0.00002108 \\
0.9  & 1 & 0.271    & 0.54241441  & 0.54172916 & 0.54160189 &  0.541557   & 0.00000131 & 0.00000131 \\
\hline
0    & 5 & 125      &   3.42245701 & $-$21.10658875 & $-$5.69164013 & $-$2.654078177 & 18.089256906  & 7.211969349 \\
0.5  & 5 & 124      &  19.24560240 &    16.12002214 &   15.00896003 &    14.25745461 &  6.325398881  & 6.027953552 \\
1    & 5 & 117      &  38.30978377 &    34.92836926 &   32.75186619 &    31.7172717  &  3.578571649  & 3.724345497 \\
2    & 5 & 109.375  &  90.38359628 &    85.34545140 &   83.48828806 &    82.7347460  &  1.08353814   & 1.096601690 \\
3    & 5 & 98       & 165.90899543 &   161.85822837 &  160.86291039 &    160.4930432 &  0.21171096   & 0.21200753 \\
4.5  & 5 & 33.875   & 339.00622801 &   338.58155153 &  338.50167840 &    338.4736317 &  0.00082310   & 0.00082305 \\
\hline
0    & 10 & 1000    &    4.49682336 & $-$2086.57335851 & $-$376.86974059 & $-$143.2486061   & 793.3606562289 & 171.83710421 \\
0.5  & 10 & 999.875 &   57.86098336 &      87.17797675 &    107.31666530 &    110.540461660 & 107.386951511  & 152.90392715 \\
1    & 10 & 999     &  163.64384127 &     277.67453375 &    254.03439794 &    238.63393904  &  80.695671484  & 113.766036855 \\
3    & 10 & 973     &  900.89151483 &     936.71840975 &    895.52388818 &    876.03124518  &  31.488571915  &  33.27162531 \\
5    & 10 & 875     & 1969.35464432 &    1925.75459775 &   1900.38182210 &    1889.89436868 &   8.27270253   &   8.321916505 \\
7.5  & 10 & 657     & 3873.03398162 &    3841.48819175 &   3834.25021073 &    3831.6076104  &   0.51510319   &   0.515184124 \\
9.5  & 10 & 142.625 & 6023.02653188 &    6021.22924375 &   6020.89380928 &    6020.7762564  &   0.00080937   &   0.0008319 \\
\end{tabular}
\end{ruledtabular}
\end{table}  
\endgroup  

Now we move to investigate $\alpha^{(1)}$ in \emph{confined} and \emph{shell-confined} plasmas by means of sample calculations on 
$1s, 2s, 2p, 3s, 3p, 3d, 4s$ states. In case of $p$ and $d$ states allowed transition happen t o final states having $\ell$-value as 
$(0,2)$ and $(1,3)$ successively. The estimated outcomes in Table~VII include contributions from both $\ell$, for same set of $R_{a}, R_{b}$
values of previous table. The third column collects the volume of ring ($V=R_{b}^{3}-R_{a}^{3}$) having inner and outer radii $R_{a},R_{b}$. 
Note that, $(R_{a},R_{b})= (0,1), (0,5), (0,10)$ explains confined plasmas. A detailed and careful analysis of this table reveals several 
fascinating features, some of them are discussed below.

\begingroup        
\squeezetable
\begin{table}
\caption{Calculated $R_{a}=R_{m}$ at ten $R_{b}$, involving $1s,2s,3s,4s$ states in \emph{shell-confined} DP and ECSCP ($\lambda_{1}, \lambda_{2}=200$ a.u) in 
field free condition.}
\centering
\begin{ruledtabular}
\begin{tabular}{l|llll|llll}
\multicolumn{9}{c}{$R_{a}=R_{m}$}        \\
\hline
  &   \multicolumn{4}{c|}{DP} &   \multicolumn{4}{c}{ECSCP} \\   
\hline 
$R_{b}$ & $1s$ & $2s$ & $3s$ & $4s$ & $1s$ & $2s$ & $3s$ & $4s$ \\
\hline
1    & 0.81776350 & 0.81844600 & 0.81857200 & 0.81861640 & 0.81776351 & 0.81844574 & 0.81857220 & 0.81861648 \\
2    & 1.38054800 & 1.38696390 & 1.38819670 & 1.38863160 & 1.38054798 & 1.38696390 & 1.38819666 & 1.38863163 \\
3    & 1.78705437 & 1.80680400 & 1.81094670 & 1.81243570 & 1.78705451 & 1.80680395 & 1.81094673 & 1.81243570 \\
4    & 2.09184539 & 2.13011530 & 2.13937730 & 2.14279860 & 2.09184635 & 2.13011505 & 2.13937718 & 2.14279852 \\
5    & 2.32953256 & 2.38598612 & 2.40275121 & 2.40915329 & 2.32953642 & 2.38598518 & 2.40275066 & 2.40915295 \\
6    & 2.52409956 & 2.59179190 & 2.61850445 & 2.62907944 & 2.52411065 & 2.59178916 & 2.61850286 & 2.62907847 \\
7    & 2.69276190 & 2.75851277 & 2.79764248 & 2.81370398 & 2.69278753 & 2.75850619 & 2.79763871 & 2.81370171 \\
8    & 2.84777754 & 2.89338024 & 2.94745556 & 2.97042156 & 2.84782806 & 2.89336637 & 2.94744777 & 2.97041687 \\
9    & 2.99755944 & 3.00126635 & 3.07296888 & 3.10435211 & 2.99764801 & 3.00123984 & 3.07295433 & 3.10434343 \\
10   & 3.14757096 & 3.08545126 & 3.17775868 & 3.21916532 & 3.14771290 & 3.08540435 & 3.17773352 & 3.21915040 \\
\end{tabular}
\end{ruledtabular}
\end{table}  
\endgroup 

\begin{enumerate}
\item
\textbf{Confined Plasmas:} In \emph{free} plasmas, $\alpha^{(1)}$ is a $(+)$ve quantity. However, in confined condition, the behavior is not so consistent, 
observing certain changes with $r_{c}$, presenting both $(-)$ve and $(+)$ve values. A candid conclusion appears as, with rise in $r_{c}$, $\alpha^{(1)}$ in a 
given $n$ with an arbitrary $\ell$ progresses as $r_{c}$ approaches towards the respective \emph{free} value. At $r_{c}=5, 10$, $2s,3s,4s$ states possesses 
$(-)$ve polarizability.
\item
\textbf{Shell-confined Plasmas:} In this environment, behavior of $\alpha^{(1)}$ changes with $\ell$. For $\ell=0$ states, (at a given $R_{b}$) it 
progresses with $R_{a}$. Similar pattern in $\alpha^{(1)}$ is also attained by altering $R_{b}$ keeping $R_{a}$ fixed. Thus, it is controls by $(R_{a}, R_{b})$ 
pair, but not by their difference, $\Delta R$. On the contrary, for $\ell \neq 0$ states, 
at a fixed $R_{b}$, it reduces with $R_{a}$. At a certain $R_{a}$, it advances with $R_{b}$. Therefore, $\alpha^{(1)}$ is dependent on all three quantities, 
$R_{a},R_{b}, \Delta R$.
\item 
\textbf{Metallic Behavior:} Through analysis of Table~VII, alludes that, for $s$-wave states, at a certain $R_{b}$, $\alpha^{(1)}$ progresses with $R_{a}$. 
But, for $2p, 3d$ states reverse trend is observed. Moreover, for $\ell=0$ states there appears a characteristic $R_{a}$ after which it prevails over 
$V$. Now, invoking the \emph{Herzfeld} criteria for \emph{insulator to metal} conversion discussed in Eq.~(\ref{eq:10})-(\ref{eq:12}), one may easily 
discern the metallic character for $s$-wave states in \emph{shell-confined} hydrogenic plasmas (both DP and ECSCP). It should be kept in mind that, no such 
feature is observed for $\ell \ne 0$ states. The threshold $R_{a}$ at which Eq.~(\ref{eq:12}) is obeyed or $\alpha^{(1)}$ becomes equal to $V$ is symbolized \
as $R_{m}$. They are reported in Table~VIII, for $\ell=0$ states at ten selected $R_{b}$ values ($1,2,3,4,5,6,7,8,9,10$). In either of the plasmas, the 
the spread of the metallic region ($R_{b}-R_{a}$) increases with progress in $R_{b}$.      
   
\end{enumerate}

\begin{figure}                         
\begin{minipage}[c]{0.50\textwidth}\centering
\includegraphics[scale=0.85]{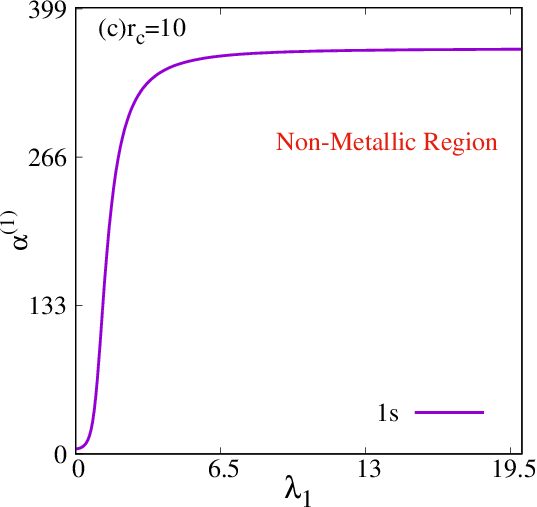}
\end{minipage}%
\begin{minipage}[c]{0.50\textwidth}\centering
\includegraphics[scale=0.85]{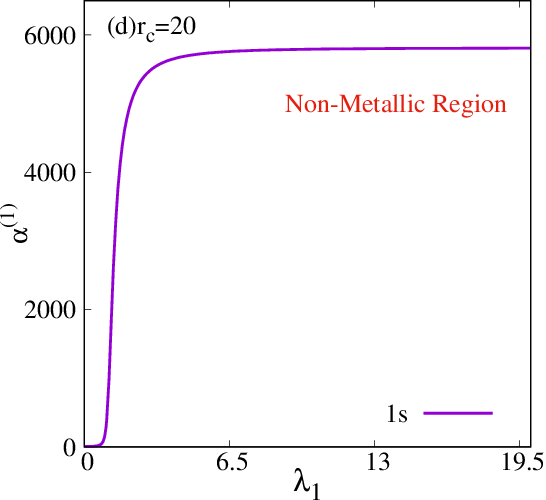}
\end{minipage}%
\vspace{3mm}
\begin{minipage}[c]{0.50\textwidth}\centering
\includegraphics[scale=0.85]{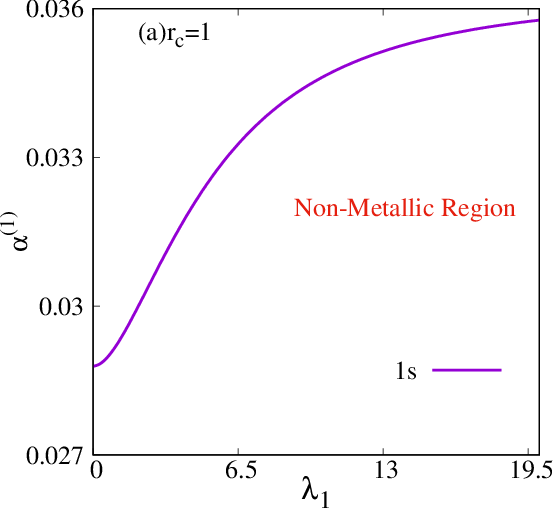}
\end{minipage}%
\begin{minipage}[c]{0.50\textwidth}\centering
\includegraphics[scale=0.85]{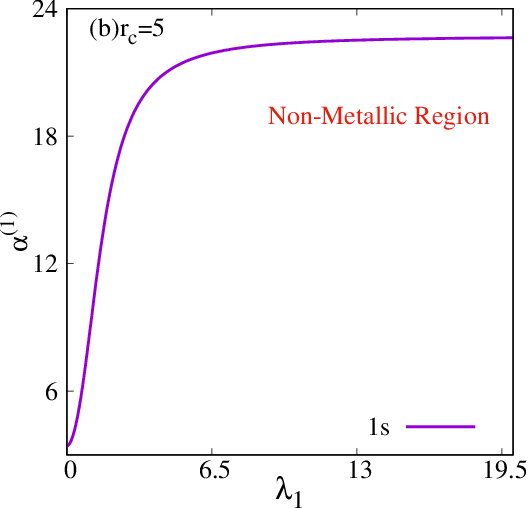}
\end{minipage}%
\caption{$\alpha^{(1)}$ values for $1s$ state in DP as a function of $\lambda_{1}$ at $r_{c}$ values $1,5,10,20$.}
\end{figure}   

Tables~VII and VIII  assert the existence of metallic character in \emph{shell-confined} DP and ECSCP. But, they are not able to ascertain the similar 
behavior in when $R_{a}=0$; i.e. in confined condition. Therefore, we have investigated $\alpha^{(1)}$ as a function of $\lambda$ in these two confined 
plasmas, keeping $r_{c}$ fixed at some arbitrary chosen values. It is useful to point out that, in relevant figures metallic $(\alpha^{(1)} > V)$ and 
non-metallic $(\alpha^{(1)}< V)$ regions are mentioned with \emph{green} and \emph{red} colors. Further, $\alpha^{(1)} = V$ is is represented by a black 
straight line.     

Figure~1 imprints the behavior $\alpha^{(1)}$ for ground state in DP with change in $\lambda_{1}$ opting $r_{c}$ fixed at four selected values, namely $1,5,10,20$. 
In all these cases it progresses with growth in $\lambda_{1}$. More importantly, $\alpha^{(1)} < V$, condition is obeyed. Therefore, metallic character is not
seen. However, for higher states with $\ell=0$, reverse pattern is seen. These has been discussed in Figure~2 and 3 involving $2s, 3s$ states.

\begin{figure}                         
\begin{minipage}[c]{0.50\textwidth}\centering
\includegraphics[scale=0.85]{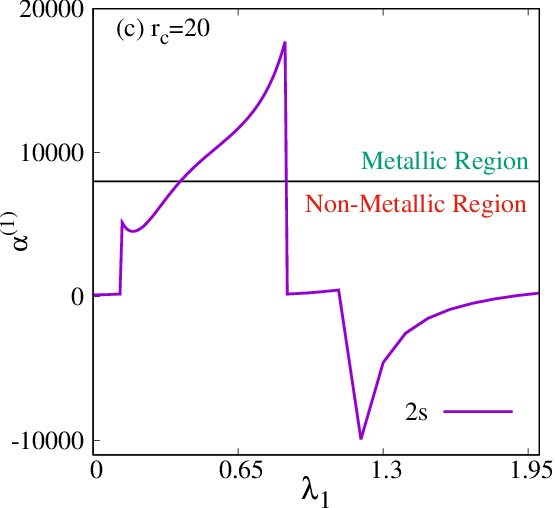}
\end{minipage}%
\begin{minipage}[c]{0.50\textwidth}\centering
\includegraphics[scale=0.85]{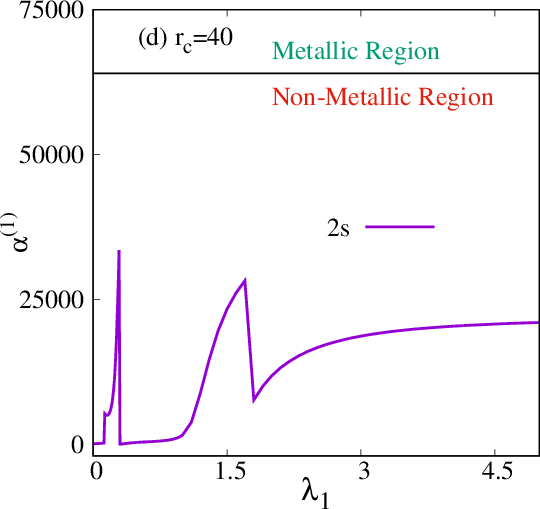}
\end{minipage}%
\vspace{3mm}
\begin{minipage}[c]{0.50\textwidth}\centering
\includegraphics[scale=0.85]{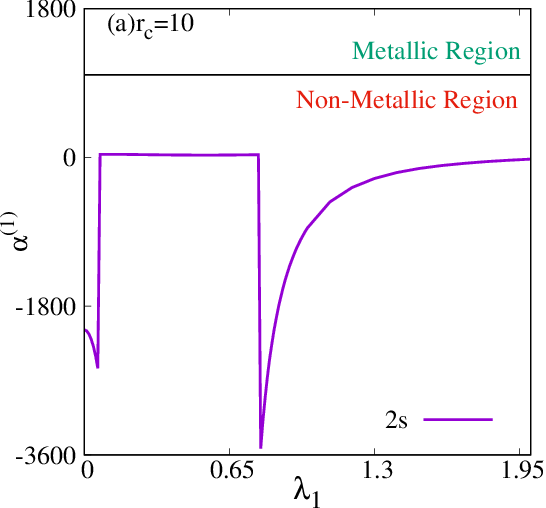}
\end{minipage}%
\begin{minipage}[c]{0.50\textwidth}\centering
\includegraphics[scale=0.85]{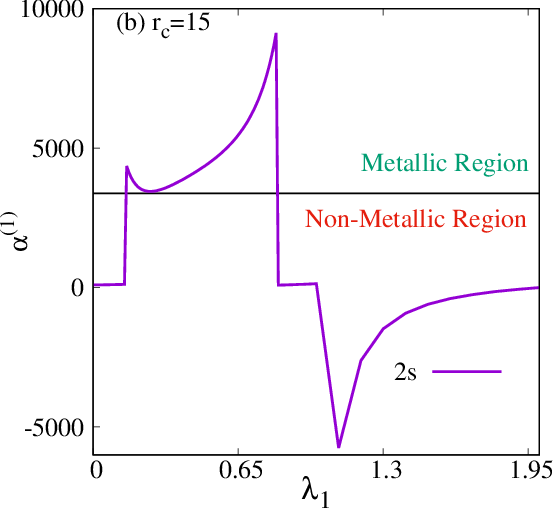}
\end{minipage}%
\caption{$\alpha^{(1)}$ values for $2s$ state in DP as a function of $\lambda_{1}$ at $r_{c}$ values $10,15,20,40$.}
\end{figure} 

Figure~2 depicts the behavior $\alpha^{(1)}$ for $2s$ state in DP with progress in $\lambda_{1}$ keeping $r_{c}$ fixed at four selected values, namely $10,15,20,40$.
In panels (a) ($r_{c}=10$) and (d) ($r_{c}=40$), $\alpha^{(1)} < V$. On the contrary, at panels  (b) ($r_{c}=15$) and (c) ($r_{c}=20$) there exists a range of 
$\lambda_{1}$, where $\alpha^{(1)} > V$. It suggests, that for $2s$ state there appears a region of $\lambda_{1}, r_{c}$, at which metallic pattern is observed. 

\begin{figure}                         
\begin{minipage}[c]{0.50\textwidth}\centering
\includegraphics[scale=0.85]{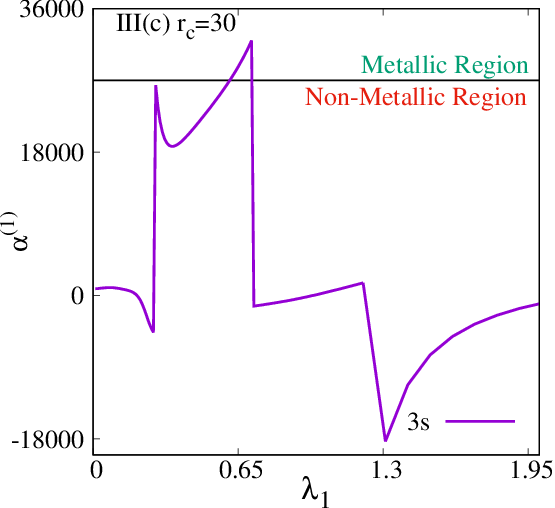}
\end{minipage}%
\begin{minipage}[c]{0.50\textwidth}\centering
\includegraphics[scale=0.85]{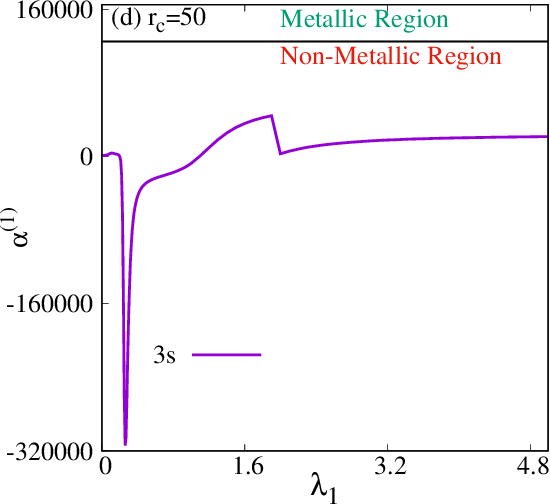}
\end{minipage}%
\vspace{3mm}
\begin{minipage}[c]{0.50\textwidth}\centering
\includegraphics[scale=0.85]{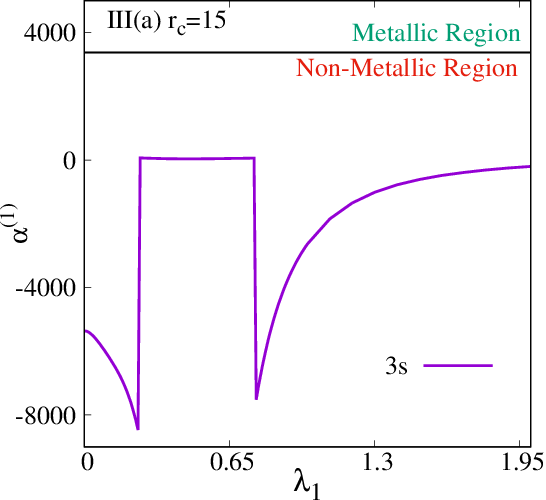}
\end{minipage}%
\begin{minipage}[c]{0.50\textwidth}\centering
\includegraphics[scale=0.85]{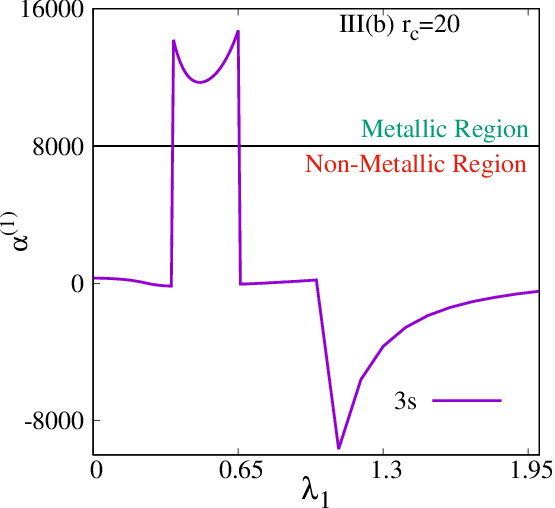}
\end{minipage}%
\caption{$\alpha^{(1)}$ values for $3s$ state in DP as a function of $\lambda_{1}$ at some selected $r_{c}$ values. See text for details.}
\end{figure} 

Figure~3 portrays the behavior $\alpha^{(1)}$ for $3s$ state in DP with growth in $\lambda_{1}$ keeping $r_{c}$ fixed at four selected values, namely $15,20,30,50$.
Similar to Fig.~2, in panels (a) ($r_{c}=15$) and (d) ($r_{c}=50$) metallic nature is not observed. However, at panels  (b) ($r_{c}=15$) and (c) ($r_{c}=20$) 
there appears a range of $\lambda_{1}$, where $\alpha^{(1)} > V$. Akin to $2s$ state, it indicates the presence of metallic like behavior in $3s$ at a certain range of 
$\lambda_{1}$. Resembling nature is seen ECSCP. In $1s$ state, $\alpha^{(1)} < V$. But, in $2s,3s$ there exists a range of $r_{c}$ and $\lambda_{2}$ at which $\alpha^{(1)} > V$. 
These, are demonstrated in Figs.~S1-S3 for $1s,2s,3s$ states respectively in supporting document. 

\begingroup              
\squeezetable
\begin{table}
\caption{$f^{(1)}$ involving $1s,2s,2p$ states for \emph{confined} DP and ECSCP for $\lambda_{1},\lambda_2=200$ a.u under external electric field 
($\cos \theta =1$ and $F$=0.1 a.u.).}
\centering
\begin{ruledtabular}
\begin{tabular}{ll|ll|ll|llll}
\multicolumn{10}{c}{DP} \\
\hline
$R_{a}$  & $R_{b}$ & $1s \rightarrow 2p$  & $1s \rightarrow 3p$ & $2s \rightarrow 2p$ & $2s \rightarrow 3p$ & 
$2p \rightarrow 1s$  & $2p \rightarrow 2s$ & $2p \rightarrow 3d$ & $2p \rightarrow 4d$ \\
\hline	
0    & 1  & 0.9847251914 & 0.0075938816 & $-$0.6085532464 & 1.5607501708 & $-$0.3282417304 & 0.2028510821 & 1.084947173 & 0.018477266 \\
0.1  & 1  & 0.8992558694 & 0.0910315700 & $-$0.5489683137 & 1.2678876762 & $-$0.2997519564 & 0.1829894379 & 1.078377462 & 0.025699465 \\
0.2  & 1  & 0.8116440851 & 0.1765877949 & $-$0.4644217903 & 1.0122462837 & $-$0.2705480283 & 0.1548072634 & 1.044551812 & 0.059606046 \\
0.5  & 1  & 0.6974624555 & 0.2896426262 & $-$0.3508436517 & 0.7344809882 & $-$0.2324874851 & 0.1169478839 & 0.929552195 & 0.173158739 \\
0.8  & 1  & 0.6699145823 & 0.3169995731 & $-$0.3234507104 & 0.6737046143 & $-$0.2233048607 & 0.1078169034 & 0.893218987 & 0.209183950 \\
\hline                                                                                            
0    & 5  & 0.8361380309 & 0.1187719538 & $-$0.4652011227 & 1.4311008380 & $-$0.2787126769 & 0.1550670409 & 1.105185469 & 0.000589816 \\
0.5  & 5  & 0.9366780393 & 0.0448308778 & $-$0.5889293266 & 1.3448509199 & $-$0.3122260131 & 0.1963097755 & 1.104946243 & 0.002554739 \\
1    & 5  & 0.8237623348 & 0.1582743437 & $-$0.4739946689 & 1.0327399751 & $-$0.2745874449 & 0.1579982229 & 1.061528816 & 0.043156235 \\
2    & 5  & 0.7205324739 & 0.2657408559 & $-$0.3731447891 & 0.7862129590 & $-$0.2401774913 & 0.1243815963 & 0.958949133 & 0.143774708 \\
3    & 5  & 0.6836004488 & 0.3033256461 & $-$0.3369872946 & 0.7036150442 & $-$0.2278668162 & 0.1123290982 & 0.911407603 & 0.191100634 \\
4.5  & 5  & 0.6673917783 & 0.3195048401 & $-$0.3209450002 & 0.6682356569 & $-$0.2224639261 & 0.1069816667 & 0.889855694 & 0.212523222 \\
\hline                                                                                            
0   & 10  & 0.7350528262 & 0.1273164444 & $-$0.3851990415 & 1.1787776832 & $-$0.2450176087 & 0.1283996805 & 1.046533203 & 0.049105808 \\
0.5 & 10  & 0.9670417912 & 0.0012262145 & $-$0.5679581943 & 1.5172373747 & $-$0.3223472637 & 0.1893193981 & 1.077766994 & 0.028556724 \\
1   & 10  & 0.8889278408 & 0.0487087750 & $-$0.4646558626 & 1.2615685051 & $-$0.2963092802 & 0.1548852875 & 1.102614229 & 0.000860195 \\
3   & 10  & 0.7466750512 & 0.2020146050 & $-$0.3576526692 & 0.8573304191 & $-$0.2488916837 & 0.1192175564 & 0.994558887 & 0.089926534 \\
5   & 10  & 0.6973476916 & 0.2823588053 & $-$0.3433822012 & 0.7345308100 & $-$0.2324492305 & 0.1144607337 & 0.929687318 & 0.168722142 \\
7.5 & 10  & 0.6720674882 & 0.3147416988 & $-$0.3254733528 & 0.6783789204 & $-$0.2240224960 & 0.1084911176 & 0.896088365 & 0.206260559 \\
9.5 & 10  & 0.6668385720 & 0.3200542481 & $-$0.3203956762 & 0.6670385229 & $-$0.2222795240 & 0.1067985587 & 0.889118103 & 0.213255692 \\
\hline
 \multicolumn{10}{c}{ECSCP} \\
\hline
$R_{a}$  & $R_{b}$ & $1s \rightarrow 2p$  & $1s \rightarrow 3p$ & $2s \rightarrow 2p$ & $2s \rightarrow 3p$ & 
$2p \rightarrow 1s$  & $2p \rightarrow 2s$ & $2p \rightarrow 3d$ & $2p \rightarrow 4d$ \\
\hline
0    & 1  & 0.984725212 & 0.007593865  & $-$0.608553283 & 1.560750223 & $-$0.328241737 & 0.202851094 & 1.0849471891 & 0.0184772537 \\
0.1  & 1  & 0.899255888 & 0.091031551  & $-$0.548968340 & 1.267887712 & $-$0.299751962 & 0.182989446 & 1.0783774765 & 0.0256994536 \\
0.2  & 1  & 0.811644092 & 0.176587787  & $-$0.464421799 & 1.012246295 & $-$0.270548030 & 0.154807266 & 1.0445518211 & 0.0596060389 \\
0.5  & 1  & 0.697462455 & 0.289642626  & $-$0.350843651 & 0.734480988 & $-$0.232487485 & 0.116947883 & 0.9295521960 & 0.1731587395 \\
0.8  & 1  & 0.669914582 & 0.316999573  & $-$0.323450710 & 0.673704615 & $-$0.223304860 & 0.107816903 & 0.8932189877 & 0.2091839509 \\
\hline                                                                                                           
0    & 5  & 0.836137691 & 0.118772104  & $-$0.465203014 & 1.431102312 & $-$0.278712563 & 0.155067671 & 1.1051853164 & 0.0005901275 \\
0.5  & 5  & 0.936678057 & 0.044830133  & $-$0.588928807 & 1.344851790 & $-$0.312226019 & 0.196309602 & 1.1049468363 & 0.0025542979 \\
1    & 5  & 0.823762447 & 0.158273646  & $-$0.473994298 & 1.032740340 & $-$0.274587482 & 0.157998099 & 1.0615293802 & 0.0431556044 \\
2    & 5  & 0.720532508 & 0.265740711  & $-$0.373144730 & 0.786213020 & $-$0.240177502 & 0.124381576 & 0.9589492016 & 0.1437745964 \\
3    & 5  & 0.683600451 & 0.303325634  & $-$0.336987289 & 0.703615049 & $-$0.227866817 & 0.112329096 & 0.9114076082 & 0.1911006258 \\
4.5  & 5  & 0.667391775 & 0.319504839  & $-$0.320945000 & 0.668235652 & $-$0.222463925 & 0.106981666 & 0.8898556952 & 0.2125232225 \\
\hline                                                                                                            
0   & 10  & 0.7350652410 & 0.127312341 & $-$0.385213057 & 1.178794763 & $-$0.245021747 & 0.128404352 & 1.0465359219 & 0.0491034056 \\
0.5 & 10  & 0.9670396465 & 0.001226961 & $-$0.567954594 & 1.517230060 & $-$0.322346548 & 0.189318198 & 1.0777697522 & 0.0285543193 \\
1   & 10  & 0.8889230773 & 0.048712046 & $-$0.464649283 & 1.261557925 & $-$0.296307692 & 0.154883094 & 1.1026139362 & 0.0008596678 \\
3   & 10  & 0.7466728691 & 0.202012739 & $-$0.357645614 & 0.857327035 & $-$0.248890956 & 0.119215204 & 0.9945564080 & 0.0899265442 \\
5   & 10  & 0.6973475211 & 0.282357651 & $-$0.343380662 & 0.734530589 & $-$0.232449173 & 0.114460220 & 0.9296871195 & 0.1687215263 \\
7.5 & 10  & 0.6720674892 & 0.314741675 & $-$0.325473330 & 0.678378921 & $-$0.224022496 & 0.108491110 & 0.8960883658 & 0.2062605435 \\
9.5 & 10  & 0.6668385642 & 0.320054248 & $-$0.320395676 & 0.667038519 & $-$0.222279521 & 0.106798558 & 0.8891181123 & 0.2132556921 \\
\end{tabular}
\end{ruledtabular}
\end{table}  
\endgroup    

\subsection{Influence of External Electric field}
In this section we discuss about the effect of external electric field on $f^{(1)}$ and $\alpha^{(1)}$. For obvious reason the numerical value will change. Here, we 
aim to discuss the change in their quantitative and qualitative pattern.

Table~IX presents estimated $f^{(1)}$ for $1s,2s,2p$ states connecting $n,\ell \rightarrow n^{'},(\ell+1)$ transitions ($n=1,2;n^{'}=2,3,4$) relating DP and ECSCP. 
Alike the field free situation, in either of the plasmas, SCC outcomes are reported for three $R_{b}$, namely, $1,5,10$; for each $R_{b}$, $R_{a}$ alters 
from $0-R_{b}$. Here, CS situation is attained when $R_{a}=0$ and $r_{c}=R_{b}=1,5,10$. Now, Comparing the results in Tables VI and IX for both DP and ECSCP, 
it is found that, for all possible transitions the qualitative pattern of $f^{(1)}$ (with increase in $R_{a}$ at a certain $R_{b}$) in $F=0$ and $F=0.1$ a.u.
are similar. Further, analysis also discloses that, in some $R_{a}, R_{b}$, the $f^{(1)}$ values in electric field are higher to respective field free cases. 
However, this behavior gets reversed in other sets of $R_{a}, R_{b}$ values.    
     
Table~X, represents the $\alpha^{(1)}$ values at $F=0.1$ a.u. involving $1s,2s,2p,3s,3d,4s$ states using same set of $R_{a}$, $R_{b}$ values of
Table~VII. In-depth comparison of Tables~VII and X for both DP and ECSCP reveals several interesting features, some of them are discussed below.

\begingroup     
\squeezetable
\begin{table}
\caption{$\alpha^{(1)}$ involving $1s,2s,3s,4s,2p,3d$ states for \emph{confined} DP and ECSCP for $\lambda_{1},\lambda_2=200$ a.u under 
external electric field ($\cos \theta =1$ and $F$=0.1 a.u.).}
\centering
\begin{tabular}{ll|l|llll|ll}
\hline
\multicolumn{9}{c}{DP} \\
\hline
$R_{a}$  & $R_{b}$ & $V$ & $\alpha^{(1)}_{1s}$ & $\alpha^{(1)}_{2s}$ & $\alpha^{(1)}_{3s}$ & $\alpha^{(1)}_{4s}$ & $\alpha^{(1)}_{2p}$ &
$\alpha^{(1)}_{3d}$  \\
\hline
0    & 1 & 1        & 0.028685 & 0.004388 & 0.001881 & 0.001051   & 0.017199 & 0.008949  \\
0.1  & 1 & 0.999    & 0.047490 & 0.027205 & 0.023578 & 0.022195   & 0.010040 & 0.008163  \\
0.2  & 1 & 0.992    & 0.072756 & 0.054885 & 0.051259 & 0.049892   & 0.005828 & 0.005659  \\
0.5  & 1 & 0.875    & 0.201853 & 0.191344 & 0.189237 & 0.188483   & 0.000831 & 0.000834  \\
0.8  & 1 & 0.488    & 0.435282 & 0.432829 & 0.432369 & 0.432208   & 0.000021 & 0.000021  \\
0.9  & 1 & 0.271    & 0.542414 & 0.541729 & 0.541601 & 0.541557   & 0.000001 & 0.000001  \\
\hline
0    & 5 & 125      &   2.226123 & $-$16.206022    &$-$6.075753  & $-$2.798394  & 14.243414 & 7.531339 \\
0.5  & 5 & 124      &  14.491464 &    17.233694    &  15.605856   &  14.624534  &  5.339967 & 6.074421 \\
1    & 5 & 117      &  32.680314 &    36.281979    &  33.559364   &  32.214396  &  3.235514 & 3.620504 \\
2    & 5 & 109.375  &  86.698183 &    86.313674    &  84.075106   &  83.097143  &  1.057095 & 1.077750 \\
3    & 5 & 98       & 164.739292 &   162.187970    & 161.061777   & 160.615365  &  0.210925 & 0.211306 \\
4.5  & 5 & 33.875   & 338.998393 &   338.583878    & 338.503064   & 338.474480  &  0.000823 & 0.000822 \\
\hline
0    & 10 & 1000    &    2.366994 & $-$48.314349 & $-$136.574119   & $-$139.154836 & 36.960919 & 54.924054 \\
0.5  & 10 & 999.875 &   19.832366 &    96.855198 &    161.913457   &    147.664876 & 16.315774 & 46.140319  \\
1    & 10 & 999     &   54.975240 &   231.979757 &    331.570336   &    295.289062 & 14.768745 & 32.917260  \\
3    & 10 & 973     &  519.371381 &  1021.276070 &    997.344383   &    938.787431 & 14.713720 & 17.336989  \\
5    & 10 & 875     & 1704.365561 &  2003.187148 &   1949.618672   &   1920.054765 &  7.088535 &  7.230665  \\
7.5  & 10 & 657     & 3842.733924 &  3850.397128 &   3839.576947   &   3834.873737 &  0.512541 &  0.512695  \\
9.5  & 10 & 142.625 & 6022.958766 &  6021.249474 &   6020.905887   &   6020.783761 &  0.000805 &  0.000827  \\
\hline
\multicolumn{9}{c}{ECSCP} \\
\hline
$R_{a}$  & $R_{b}$ & $V$ & $\alpha^{(1)}_{1s}$ & $\alpha^{(1)}_{2s}$ & $\alpha^{(1)}_{3s}$ & $\alpha^{(1)}_{4s}$ & $\alpha^{(1)}_{2p}$ &
$\alpha^{(1)}_{3d}$  \\
\hline
0    & 1 & 1        & 0.028685 & 0.004388 & 0.001881 & 0.001051 & 0.017199 & 0.008949 \\
0.1  & 1 & 0.999    & 0.047490 & 0.027205 & 0.023578 & 0.022195 & 0.010040 & 0.008163 \\
0.2  & 1 & 0.992    & 0.072756 & 0.054885 & 0.051259 & 0.049892 & 0.005828 & 0.005659 \\
0.5  & 1 & 0.875    & 0.201853 & 0.191344 & 0.189237 & 0.188483 & 0.000831 & 0.000834 \\
0.8  & 1 & 0.488    & 0.435282 & 0.432829 & 0.432369 & 0.432208 & 0.000021 & 0.000021 \\
0.9  & 1 & 0.271    & 0.542414 & 0.541729 & 0.541601 & 0.541557 & 0.000001 & 0.000001 \\
\hline
0    & 5 & 125      &   2.226022 & $-$16.205347    & $-$6.075795 & $-$2.798411 & 14.242942 & 7.531355 \\
0.5  & 5 & 124      &  14.490991  &   17.233831    &   15.605931 &   14.624579 &  5.339841 & 6.074408 \\
1    & 5 & 117      &  32.679698  &   36.282142    &   33.559464 &   32.214457 &  3.235469 & 3.620484 \\
2    & 5 & 109.375  &  86.697749  &   86.313791    &   84.075178 &   83.097187 &  1.057092 & 1.077747 \\
3    & 5 & 98       & 164.739152  &  162.188009    &  161.061801 &  160.615380 &  0.210925 & 0.211306 \\
4.5  & 5 & 33.875   & 338.998394  &  338.583881    &  338.503067 &  338.474481 &  0.000823 & 0.000822 \\
\hline
0    & 10 & 1000    &    2.366863 &$-$48.305017   & $-$136.552211 & $-$139.150610 & 36.955529 & 54.916435 \\
0.5  & 10 & 999.875 &   19.830950 &   96.847711   &    161.914596 &    147.669598 & 16.313636 & 46.133709 \\
1    & 10 & 999     &   54.971399 &  231.964474   &    331.573761 &    295.296156 & 14.766828 & 32.912464 \\
3    & 10 & 973     &  519.346824 & 1021.271740   &    997.355997 &    938.795020 & 14.712319 & 17.335371 \\
5    & 10 & 875     & 1704.338447 & 2003.194993   &   1949.624430 &   1920.058295 &  7.088361 &  7.230499 \\
7.5  & 10 & 657     & 3842.730471 & 3850.398145   &   3839.577555 &   3834.874125 &  0.512541 &  0.512695 \\
9.5  & 10 & 142.625 & 6022.958829 & 6021.249507   &   6020.906043 &   6020.783663 &  0.000762 &  0.000842 \\
\hline
\end{tabular}
\end{table}  
\endgroup 

\begin{enumerate}
\item 
For $1s,2p,3d$ states, $\alpha^{(1)}$ in presence of external field possesses higher value compare to its free counterpart. However, 
for $2s,3s,4s$ states behavior is not very consistent. 

\item
Similar to field free cases, here also $\alpha^{(1)}$ can have both $-$ve and $-$ve values. Particularly, for $2s,3s,4s$ state such 
phenomena is observed.

\item 
For $1s,2s,3s,4s$ states metallic behavior is observed after a threshold value of $R_{a}$ (at a fixed $R_{b}$). 

\item 
The threshold $R_{a}$ at which Eq.~(\ref{eq:12}) is valid are reported in Table~S1 in supporting document, for $1s,2s,3s,4s$ states at ten selected $R_b$ 
values ($1, 2, 3, 4, 5, 6, 7, 8, 9, 10$). As usual, in either of the plasmas, the spread of the metallic region ($R_{b}-R_{a}$) 
increases with growth in $R_{b}$.
  
\end{enumerate}

\begin{figure}                         
\begin{minipage}[c]{0.5\textwidth}\centering
\includegraphics[scale=0.85]{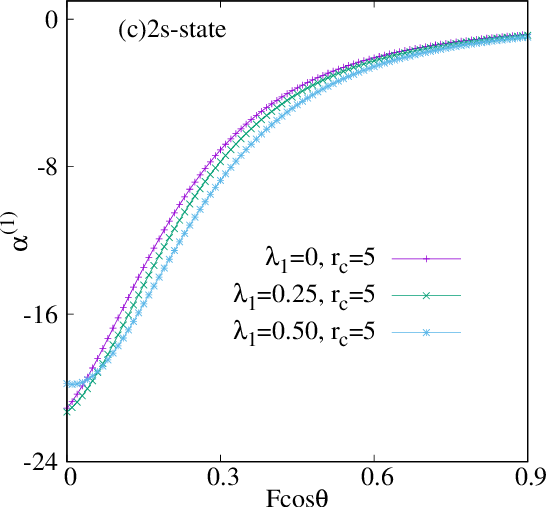}
\end{minipage}%
\begin{minipage}[c]{0.5\textwidth}\centering
\includegraphics[scale=0.85]{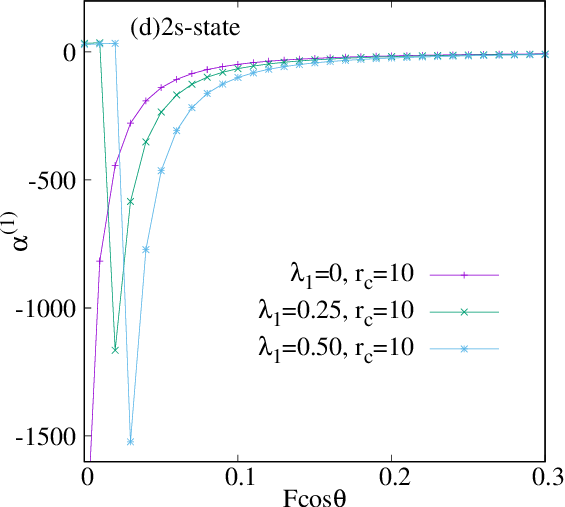}
\end{minipage}%
\vspace{3mm}
\begin{minipage}[c]{0.5\textwidth}\centering
\includegraphics[scale=0.86]{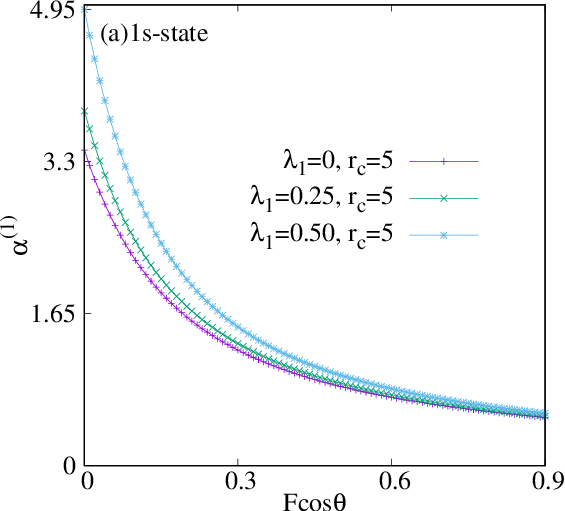}
\end{minipage}%
\begin{minipage}[c]{0.5\textwidth}\centering
\includegraphics[scale=0.86]{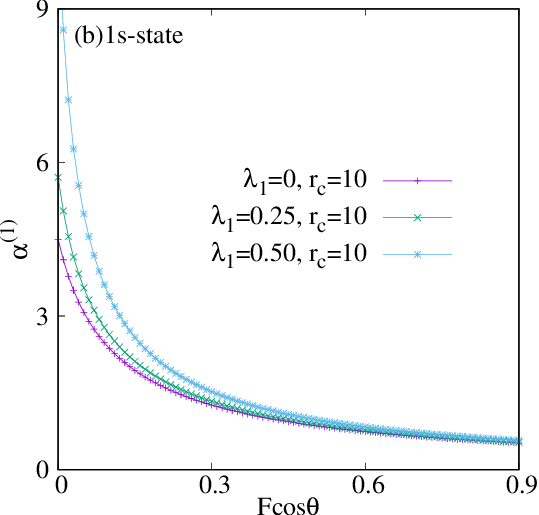}
\end{minipage}%
\caption{$\alpha^{(1)}$ values for $1s, 2s$ states in DP as a function of $F\cos\theta$ at some selected $\lambda_{1}$ values.}
\end{figure} 

Now, to move to investigate the effect of field strength $F\cos\theta$ on $\alpha^{(1)}$. In this context, pilot calculations are done at 
six separate sets of $(\lambda, r_{c})$, they are $(0,5),(0,10),(0.25,5),(0.25,10),(0.5,5),(0.5,10)$. Panels~(a)-(b) in Fig.~4, imprints 
the change in $\alpha^{(1)}$ values with advancement of $F\cos\theta$ in $1s$ state of DP at three different $\lambda_{1}$ values keeping $r_{c}$ 
fixed at $5$ and $10$ respectively. At a certain $\lambda_{1},r_{c}$ it decreases with increase in field strength. However, at a fixed $r_{c}$, 
it increases with increase in $\lambda_{1}$. Following the same line, $\alpha^{(1)}$ in $2s$ state is plotted in panels~(c)-(d) in Fig.~(4) at the 
same six sets of ${\lambda, r_{c}}$. It is interesting to point out that, $\alpha^{(1)}$ behaves differently with at $r_{c}=5$ and $10$ successively. 
At $r_{c}=5$ (panel (c)), $\alpha^{(1)}$increases with progress in $F\cos\theta$. Further, it possesses $-$ve value through the range of field 
strength. Panel (d) ($r_{c}=10$), shows that, at $\lambda_{1}=0$, starting from initial $-$ve value it increases with rise in $F\cos\theta$. However, 
at $\lambda_{1}=0.25, 0.50$, it has $+$ve value at the beginning. Then sharply reduces to $-$ve value. Finally, after reaching the minimum 
it increases but never becomes $-$ve.                  

\begin{figure}                         
\begin{minipage}[c]{0.5\textwidth}\centering
\includegraphics[scale=0.85]{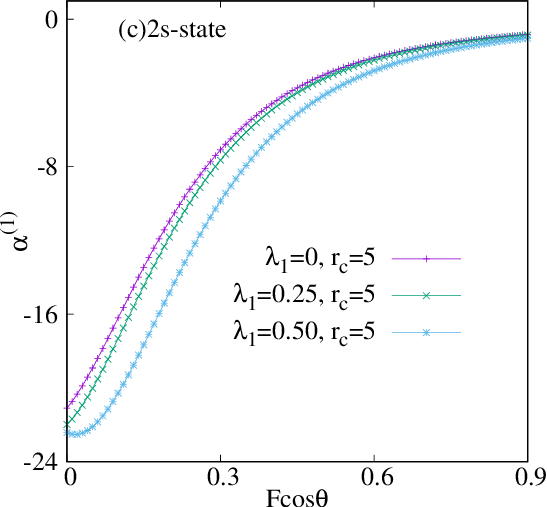}
\end{minipage}%
\begin{minipage}[c]{0.5\textwidth}\centering
\includegraphics[scale=0.85]{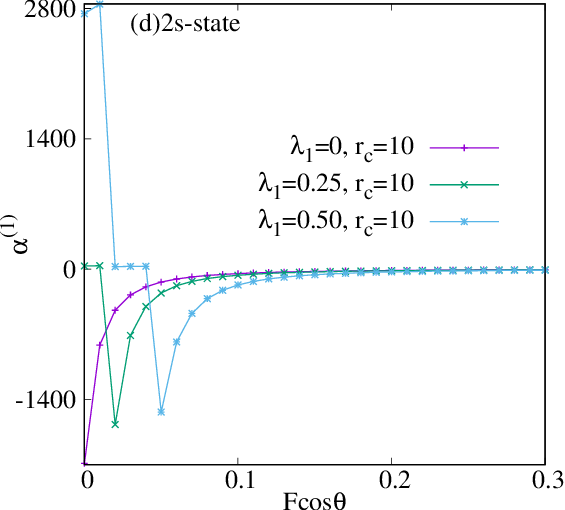}
\end{minipage}%
\vspace{3mm}
\begin{minipage}[c]{0.5\textwidth}\centering
\includegraphics[scale=0.86]{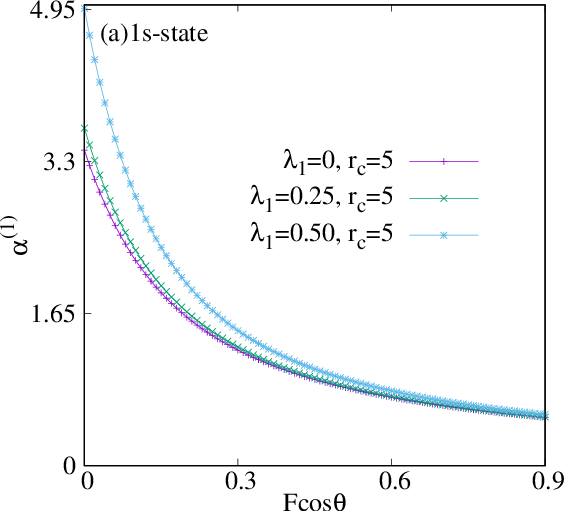}
\end{minipage}%
\begin{minipage}[c]{0.5\textwidth}\centering
\includegraphics[scale=0.86]{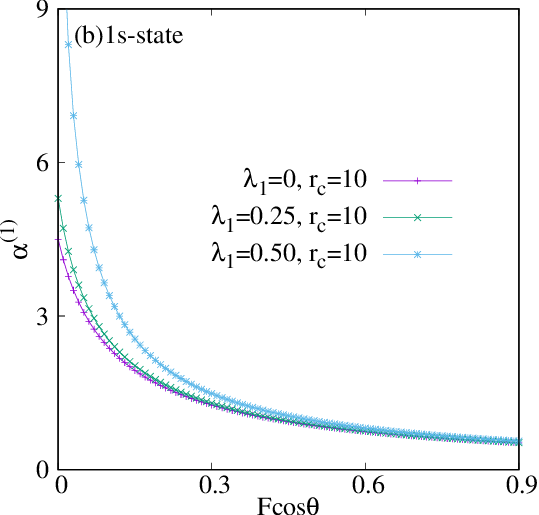}
\end{minipage}%
\caption{$\alpha^{(1)}$ involving $1s, 2s$ states in ECSCP as a function of $F\cos\theta$ at some selected $\lambda_{2}$ values.}
\end{figure} 

Figure~5, reveals the behavior of $\alpha^{(1)}$ in ECSCP with increase in field strength. Panels (a)-(b) portrays the outcome for $1s$ state at same 
six sets of ${\lambda, r_{c}}$ values. The ground state of ECSCP exhibit exactly similar pattern to DP with obvious changes in their numerical values. Moreover, 
in $2s$ state at $r_{c}=5$ (panel (c)) we observe identical nature. But, at $r_{c}=10$ (panels (d)) distinguished shift is observed at $\lambda_{2}=0.25,0.50$.
In these cases, with rise in field strength, $\alpha^{(1)}$ sharply drops closes to \emph{zero} to reach a flat plateau then again exquisitely falls to a
minimum and finally increases.    
  
\begin{figure}                         
\begin{minipage}[c]{0.5\textwidth}\centering
\includegraphics[scale=0.86]{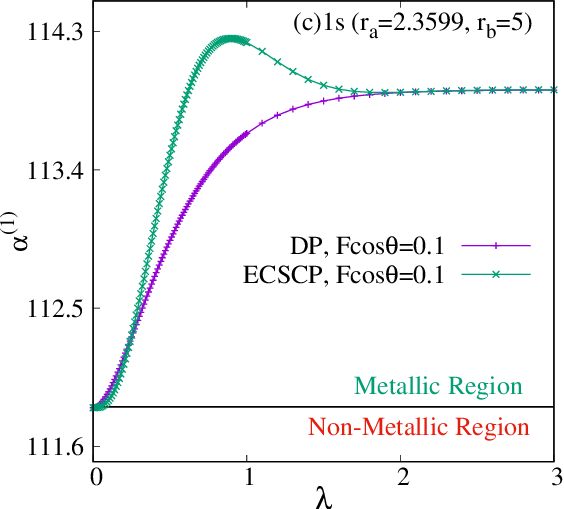}
\end{minipage}%
\begin{minipage}[c]{0.5\textwidth}\centering
\includegraphics[scale=0.86]{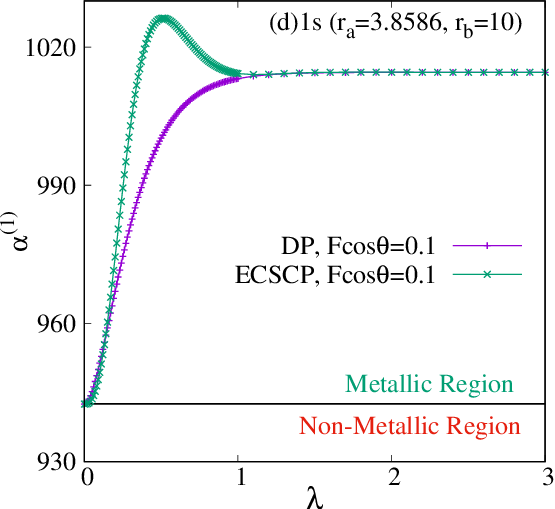}
\end{minipage}%
\vspace{3mm}
\begin{minipage}[c]{0.5\textwidth}\centering
\includegraphics[scale=0.86]{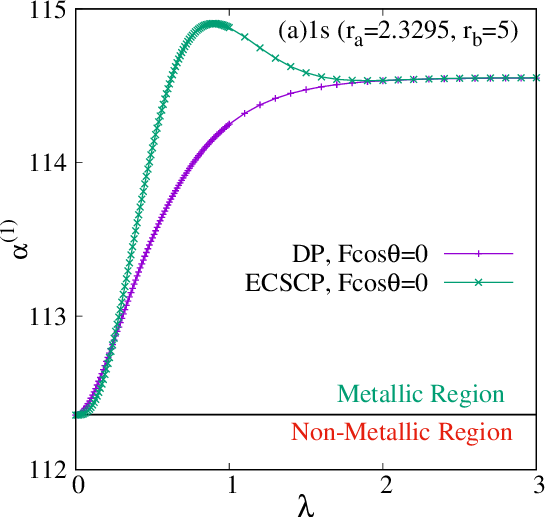}
\end{minipage}%
\begin{minipage}[c]{0.5\textwidth}\centering
\includegraphics[scale=0.86]{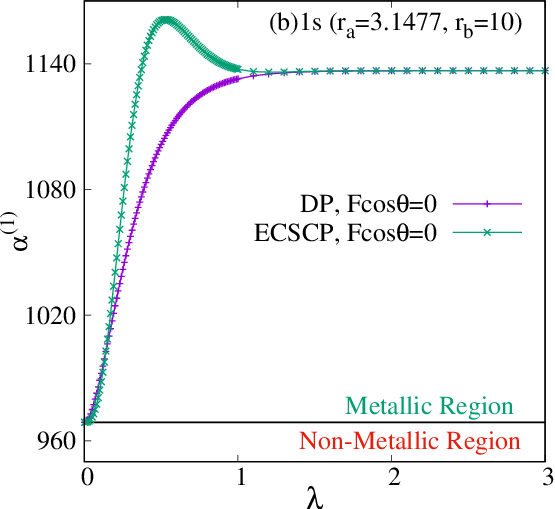}
\end{minipage}%
\caption{$\alpha^{(1)}$ values for $1s$ state in DP, ECSCP as a function of $\lambda$ at two different $r_{b}$ ($5,10$ a.u) values choosing $F\cos\theta =0$ and 
$F\cos\theta=0.1 a.u.$.}
\end{figure}

Now, we probe the influence of $\lambda$ on the metallic behavior of \emph{shell-confined} DP and ECSCP in $F\cos\theta=0$ and $F\cos\theta=0.1$ a.u. at two different 
$R_{b}$ namely, $5,10$ in $1s, 2s$ states. Here, $R_{a}$ has been opted as the respective $R_{m}$ values. The bottom panels (a)-(b) of Fig.~6 demonstrates the 
results in $1s$ state for $r_{c}=5,10$ respectively under field free conditions. In both DP and ECSCP $\alpha^{(1)} > V$ through out the range of $\lambda$. For 
both $R_{b}$, $\alpha^{(1)}$ increases with increase in $\lambda_{1}$. However, in ECSCP, in either of $R_{b}$, it advances to a maximum then approaches to 
respective DP limits. Further, the top panels (c)-(d) exhibit the same for $F\cos \theta=0.1$ a.u. In this case identical behavioral pattern to field free condition 
is observed for both DP and ECSCP. In essence, it can be said that, in both the plasmas metallic character in ground state increases with increase in $\lambda$.            

\begin{figure}                         
\begin{minipage}[c]{0.5\textwidth}\centering
\includegraphics[scale=0.86]{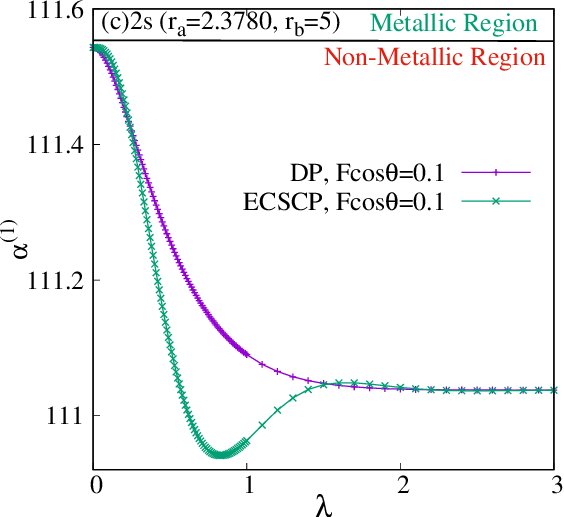}
\end{minipage}%
\begin{minipage}[c]{0.5\textwidth}\centering
\includegraphics[scale=0.86]{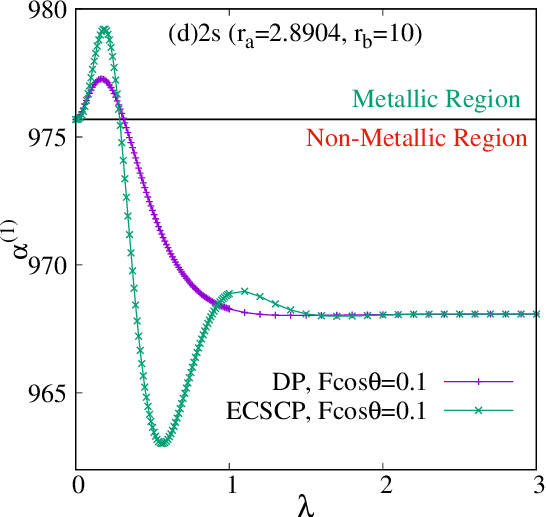}
\end{minipage}%
\vspace{3mm}
\begin{minipage}[c]{0.5\textwidth}\centering
\includegraphics[scale=0.86]{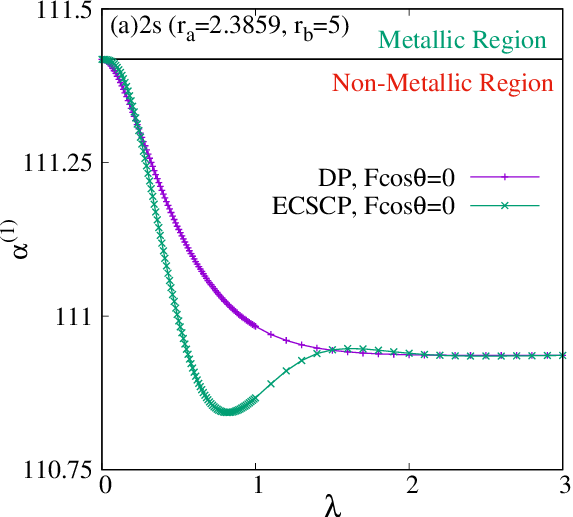}
\end{minipage}%
\begin{minipage}[c]{0.5\textwidth}\centering
\includegraphics[scale=0.86]{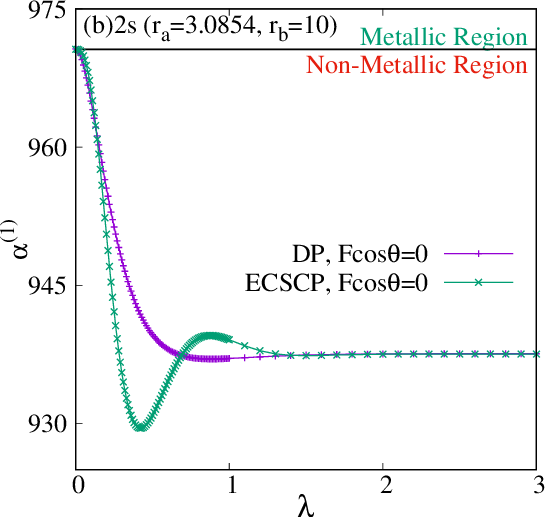}
\end{minipage}%
\caption{$\alpha^{(1)}$ values for $2s$ state in DP, ECSCP as a function of $\lambda$ at two different $r_{b}$ ($5,10$ a.u) values choosing $F\cos\theta =0$ and 
$F\cos\theta=0.1 a.u.$.}
\end{figure} 

The panels (a)-(c) in Fig.~7 explains the change in $\alpha^{(1)}$ with rise in $\lambda$ in $2s$ state of \emph{shell-confined} DP and ECSCP at $R_{b}=5$ involving 
$F\cos \theta=0$, $F\cos \theta=0.1$ a.u successively. It is necessary to mention that, at the onset $\alpha^{(1)} > V$, but with rise in $\lambda$, we observe reverse 
scenario $(\alpha^{(1)} < V)$. In DP involving both $F\cos \theta=0$, $F\cos \theta=0.1$ a.u cases, $\alpha^{(1)}$ decreases with rise in $\lambda_{1}$. However, in ECSCP 
in either of the conditions, it decreases to reach a minimum then increases to attain a shallow maximum before merging to respective DP values. Panel (c)-(d) indicate the 
changes of $\alpha^{(1)}$ with and with out field respectively at $R_{b}=10$. As usual the initial $\alpha^{(1)}$ values are higher than $V$. In field free DP, it decreases 
with progress in $\lambda_{1}$. However, under the external electric field it increase to reach a maximum then reduces. In ECSCP, relating filed free case, it decreases to 
approach a minimum then increases to a prominent maximum before reaching to DP values. At $F\cos \theta=0.1$ a.u, it climbs a maximum then falls down to a minimum and again 
attains a maximum to convene to DP limit.          

\section{future and outlook}
Degeneracies, dipole oscillator strength and polarizability have been investigated for confined DP and ECSCP applying GCS model with special emphasis on 
\emph{shell-confined} condition, which has not been done before. This model can illustrate both \emph{confined} and \emph{free} conditions efficiently. Effect of 
external electric field on these properties is also elaborately investigated. Impact of field strength on these spectroscopic properties is also probed. An in-depth analysis 
reveals several impressive and hitherto unreported characteristics in these plasmas. Existance of an additional degeneracy in weakly coupled 
plasmas under confined environment has also been established. In GCS with growth in $n$, count of these incidental degenerate states increases, conversely, at a fixed $n$, with rise in $\ell$, their 
number decreases. In strressed condition, \emph{negative} polarizability is experienced in excited states. Further, metallic nature
is observed in both \emph{confined} and \emph{shell-confined} conditions. The influence of $R_{a}, R_{b}$ on spectroscopic properties are probed. Similar calculations in other central 
potentials and plasma systems is highly required. Especially, it is desirable to justify the existence of these degeneracies on 
other experimental plasmas. Investigation of Hellmann-Feynman theorem for confined plasmas is necessary. Further, investigation of photo-ionization cross-section, relative information, 
two-photon transition amplitude,  in \emph{confined} and \emph{shell-confined} plasmas would provide vital insight. Moreover, present study can be extended to many-electron atomic plasmas.              

\section{Acknowledgement}
NM thanks CSIR, New Delhi, India, for a Senior Research Associate-ship (Pool No. 9033A). The author acknowledges Prof. A. K. Roy for laboratory support and several valuable 
comments. 


\end{document}